\documentclass[12pt,preprint]{aastex}

\newcommand{\ee}[1]{\mbox{${} \times 10^{#1}$}}

\newcommand{\h}{\mbox{$^h$}}
\newcommand{\m}{\mbox{$^m$}}

\newcommand{\degree}{\mbox{$^{\circ}$}}
\newcommand{\ptdeg}{\mbox{$^{\circ}$}}

\newcommand\cmv{\mbox{cm$^{-3}$}}

\newcommand{\spitzer}{\mbox{\it Spitzer}}

\newcommand{\msun}{\mbox{M$_\odot$}}

\newcommand{\td}{\mbox{$T_D$}}

\newcommand{\mean}[1]{\mbox{$\langle#1\rangle$}} 
\newcommand{\av}{\mbox{$A_V$}} 

\newcommand{\chisq}{\mbox{$\chi_r^2$}}
\newcommand{\beam}{\mbox{$\theta_{mb}$}}

\newcommand{\irassrc}{\mbox{IRAS 16293$-$2422}}

\begin{document}

\title {\bf Bolocam Survey for 1.1 mm Dust Continuum Emission in the c2d Legacy Clouds. II. Ophiuchus}
\author{Kaisa E. Young}
\affil{Department of Astronomy, The University of Texas at Austin,
       1 University Station C1400, Austin, Texas 78712--0259}
\author{Melissa L. Enoch}
\affil{Division of Physics, Mathematics \& Astronomy, California Institute of 
Technology, Pasadena, CA 91125}
\author{Neal J. Evans II}
\affil{Department of Astronomy, The University of Texas at Austin,
       1 University Station C1400, Austin, Texas 78712--0259}
\email{nje@astro.as.utexas.edu}
\author{Jason Glenn}
\affil{Center for Astrophysics and Space Astronomy, 389-UCB, University of 
Colorado, Boulder, CO 80309}
\email{Jason.Glenn@colorado.edu}
\author{Anneila Sargent}
\affil{Division of Physics, Mathematics \& Astronomy, California Institute of 
Technology, Pasadena, CA 91125}
\email{afs@phobos.caltech.edu}
\author{Tracy L. Huard}
\affil{Harvard-Smithsonian Center for Astrophysics, 60 Garden Street,
Cambridge, MA 02138}
\email{thuard@cfa.harvard.edu}
\author{James Aguirre\altaffilmark{1}}
\affil{Jansky Fellow, National Radio Astronomy Observatory}
\altaffiltext{1}{Center for Astrophysics and Space Astronomy, 389-UCB, 
University of Colorado, Boulder, CO 80309}
\email{James.Aguirre@Colorado.EDU}
\author{Sunil Golwala}
\affil{Division of Physics, Mathematics \& Astronomy, California Institute of 
Technology, Pasadena, CA 91125}
\email{golwala@phobos.caltech.edu}
\author{Douglas Haig}
\affil{ Physics and Astronomy, Cardiff University, 5, The Parade, P.O. 
Box 913, Cardiff, CF24 3YB Wales, UK}
\email{Douglas.Haig@astro.cf.ac.uk}
\author{Paul Harvey}
\affil{Department of Astronomy, The University of Texas at Austin,
       1 University Station C1400, Austin, Texas 78712--0259}
\email{pmh@astro.as.utexas.edu}
\author{Glenn Laurent}
\affil{Center for Astrophysics and Space Astronomy, 389-UCB, University of 
Colorado, Boulder, CO 80309}
\email{Glenn.Laurent@Colorado.EDU}
\author{Philip Mauskopf}
\affil{ Physics and Astronomy, Cardiff University, 5, The Parade, P.O. 
Box 913, Cardiff, CF24 3YB Wales, UK}
\email{Philip.Mauskopf@astro.cf.ac.uk}
\author{Jack Sayers}
\affil{Division of Physics, Mathematics \& Astronomy, California Institute of 
Technology, Pasadena, CA 91125}
\email{jack@its.caltech.edu}

\begin{abstract}


We present a large-scale millimeter continuum map of the
Ophiuchus molecular cloud. Nearly 11 square degrees, including
all of the area in the cloud with $A_V \ge 3$ magnitudes,  was mapped
at 1.1 mm with Bolocam on the Caltech Submillimeter Observatory
(CSO). By design,
the map also covers the region mapped in the infrared
with the {\it Spitzer Space Telescope}.
We detect 44 definite sources, and a few likely sources
are also seen along a filament in the eastern streamer.
The map indicates that dense cores in Ophiuchus are very clustered
and often found in filaments within the cloud.  
Most sources are round, as measured at the half
power point, but elongated when measured at lower contour levels, suggesting
spherical sources lying within filaments.
The masses, for an assumed dust temperature of 10~K, range from 0.24 to
3.9 \msun, with a mean value of 0.96 \msun. The  total mass in
distinct cores is 42 \msun, 0.5 to 2\% of the total cloud mass, and  
the total mass above 4 $\sigma$ is about 80 \msun.
The mean densities in the cores are quite high,
with an average of $1.6\ee6$ \cmv, suggesting short free-fall times.
The core mass distribution can be fitted with a power law with slope 
$\alpha=2.1 \pm 0.3$ for $M>0.5$ \msun, similar to that found in other
regions, but slightly shallower than that of some determinations of the
local IMF.  In agreement with previous studies, our survey shows that dense
cores account for a very small fraction of the cloud volume and total mass.
They are nearly all confined to regions with $\av \geq 9$ mag, a lower
threshold than found previously.

\end{abstract}

\keywords{ISM: dust --- ISM: submillimeter}

\section{Introduction}

For over 30 years, astronomers have known that stars are born in
molecular clouds. However, the fraction of cloud mass that forms stars
is usually small \citep{leisawitz89}, and the crucial step seems to be 
the formation of
dense cores, which are well traced by dust continuum emission.
Understanding the processes that control the formation of dense cores
within the molecular cloud is necessary for understanding the efficiency
and distribution of star formation \citep{evans99}. 
Progress in this area requires complete
maps of large molecular clouds in the millimeter continuum emission, which 
traces the location, mass, and other properties of the dense cores.
Such maps are becoming feasible only with the arrival of millimeter-wave
cameras such as Bolocam.

One well-known
birthplace of stars is the Ophiuchus molecular cloud. Located at a
distance of $125\pm25$ pc (\citealt{deg89}), Ophiuchus contains the L1688
dark cloud region, which contains the Ophiuchus cluster 
(16\h\ 27\m, $-$24\degree\ 30\arcmin\ (J2000)) of young stars and embedded
objects. The cluster region
has been studied in great detail at a variety of wavelengths from
millimeter molecular lines (\citealt{lor89}; \citealt{ridge06})
to near-infrared 
(e.g., \citealt{wilking89}; \citealt{all02}) 
to X-ray \citep{ima01}. It has also been mapped in dust continuum
emission (\citealt{joh00}; \citealt{mot98}). 
The embedded cluster is itself surrounded by a somewhat older
population of stars extending over 1.3 deg$^2$ \citep{wilking05}. 
The cloud is home to two other known
regions of star formation, the Lynds dark clouds L1689 (16\h\ 32\m,
$-$24\degree\ 29\arcmin) and L1709 (16\h\ 31\m, $-$24\degree\
03\arcmin). However, little is known about star formation outside of
these three regions.

In this paper, we present the first large-scale millimeter continuum map of the
entire Ophiuchus molecular cloud.  Maps at millimeter wavelengths of the
dust continuum emission find regions of dense gas and dust, both those
with embedded protostars and those that are starless.
Previously published maps of the Ophiuchus cloud covered only
small regions: \citet{mot98} mapped about 0.13 deg$^2$ at 1.3 mm and
\citet{joh00} mapped about 0.19 deg$^2$ at 850 \micron.  A larger
(4 deg$^2$) map at 850 \micron\ is referred to but not published 
by \citet{joh04}; it will be shown in \citet{ridge06}. 
Most recently, \citet{stanke05} have mapped a
1.3 deg$^2$ area of L1688 at 1.2 mm.
Our map covers 10.8 deg$^2$, providing a total picture of
the dense gas in Ophiuchus. 

\begin{figure}
\plotone{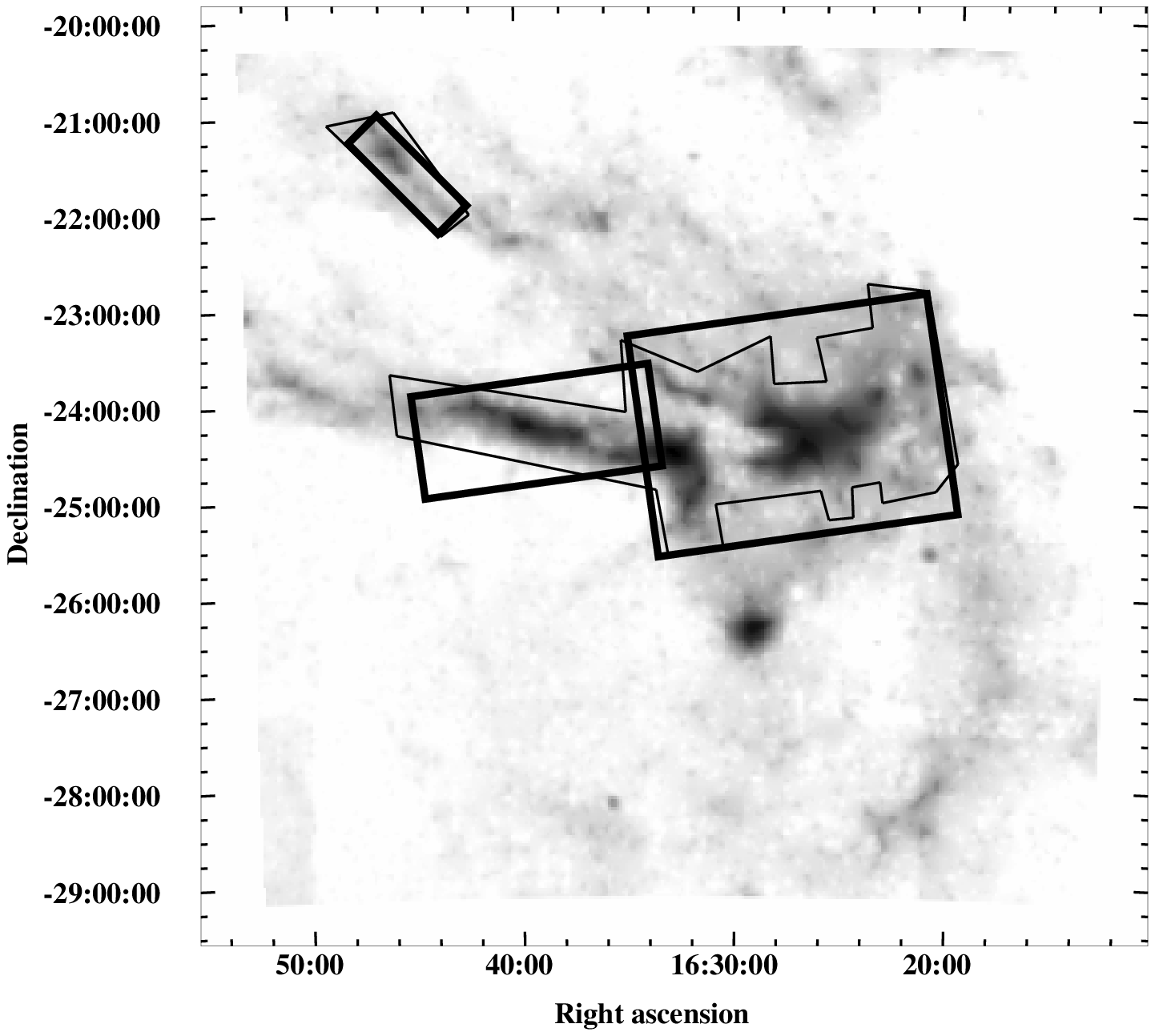} \figcaption{\label{obs} Extinction map of
Ophiuchus from Cambr\'{e}sy (1999) with the outline of the Bolocam
observation area (thick lines) and the \spitzer\ IRAC observation area
(thin lines).  
The area observed with IRAC was chosen to cover the cloud 
down to $A_V \ge 3$.  
The Bolocam observations were designed to cover approximately the 
same region observed with IRAC.}
\end{figure}

Our survey complements the \spitzer\ c2d Legacy project
``From Molecular Cores to Planet-forming Disks'' \citep{eva03}.   
All of the area in the cloud with $A_V
\ge 3$ magnitudes (according to the map of \citet{cam99}) was observed
with Bolocam and the InfraRed Array Camera (IRAC) 
on \spitzer\ (Figure \ref{obs}).
A somewhat larger area was also mapped at 24, 70, and 160 \micron\ with the
Multiband Imaging Photometer for Spitzer (MIPS).
Maps of millimeter molecular line emission for this
same area have been made by the COMPLETE team\footnote{see 
http://cfa-www.harvard.edu/COMPLETE} (\citealt{goo04}; \citealt{ridge06}). 
Previously, the largest maps
of molecular lines were those of \citet{lor89} and \citet{tac00}.

In addition to Ophiuchus, two other
clouds were mapped with both Bolocam and \spitzer\ by the c2d team,
Perseus (Enoch et al.\ 2006, hereafter Paper~I) and Serpens 
(Enoch et al.\ 2006, in prep.). 
This paper is the second in a series describing these
observations and applies the analysis methods described in Paper~I.
The results for Ophiuchus will be compared to those for Perseus in
\S \ref{compsect}.






\section{Observations}

We mapped the Ophiuchus molecular cloud at 1.12 mm (hereafter 1.1 mm
for brevity) with Bolocam on
the CSO\footnote{The CSO is operated by the California Institute of
Technology under funding from the National Science Foundation,
contract AST-0229008.} during two observing runs: 21 May -- 09 June
2003 and 06 -- 11 May 2004.  
Bolocam is a 144-element bolometer array camera that operates
at millimeter wavelengths \citep{gle03}. In May 2003, there were 95 channels;  
in May 2004, the observations were taken with 114 channels.  
The Bolocam field of view is $7\farcm5$, the beams are well described by a 
Gaussian with a FWHM of 31\arcsec\ at 1.1~mm, and the 
instrument has a bandwidth of 45~GHz. 
The cloud was observed in three main sections: the main part,
including the L1688 cluster region, the large eastern streamer that
extends to the east of L1689, and a smaller northeastern streamer. The
map of the northeastern streamer is not contiguous with the other regions, as
shown in Figure \ref{obs}.

Each section was observed by scanning Bolocam at a rate of 60\arcsec\
per second without chopping. Each subsequent subscan was offset from the previous one by 162\arcsec\
perpendicular to the scan direction. With this scan pattern, 1 square
degree was observed with 23 subscans in approximately half an hour
of telescope time, including 20-second turn-around times at the edges of the maps.

Each section of the map was scanned in two orthogonal directions, rotated
slightly from right ascension (RA) and declination (Dec) by
small angles. This technique allows for good cross-linking of the
final map with sub-Nyquist sampling and minimal striping from $1/f$ noise.  
The northeastern
streamer is a little more than 
0.5 deg$^2$, the eastern streamer section is about 2.7 deg$^2$,
and the large L1688/main
cloud section covers a total of 7.4 deg$^2$, which was observed in four
sections of approximately 4 deg$^2$ each.

The best-weather observations from both runs for each of the three sections
were averaged and combined into a single large map:
six observations of the northeastern streamer were averaged, three in RA 
and three in Dec, for a total observation time
of about 4.4 hours. The eastern streamer sections 
consists of five observations (three RA and two  Dec scans) for a total observation 
time of 6 hours. The main cloud region was observed in four pieces, each observed four times (two RA and two Dec scans), which required thirteen hours of integration time.  The resulting coverage varies by $\sim25$\%.

In addition to the maps of Ophiuchus, small maps of secondary calibrators and 
pointing sources were made every 2 hours throughout the run.  
All calibration sources observed throughout each run were used to 
derive the flux calibration factor for that run.
Planets provided beam maps and 
primary flux calibration sources. Uranus and Mars were observed during both 
runs. Neptune was also observed on the May 2004 run.

\section{Data Reduction}\label{data}   

\subsection{Pointing and Flux Calibration}\label{pointing}

Beam maps were made from observations of Uranus, Mars, and Neptune, and a 
pointing model was generated with observations of these planets, 
Galactic HII regions, and the protostellar source \irassrc, 
which lies in the Ophiuchus cloud complex.  
For May 2004, the root-mean-square (rms)
pointing uncertainty derived from the deviations of \irassrc\
centroids from the Bolocam/CSO pointing model is 2 to 3\arcsec;  however,
the number of \irassrc\ observations is small (seven), so the rms is
not well-characterized.
For May 2003, the rms pointing uncertainty was 6\arcsec, based on the 
dispersion of the centroid of \irassrc\ 
after the pointing model was applied.

A flux density calibration curve was generated for each run from observations of Uranus, Neptune, and secondary calibrators (Sandell 1994) throughout each night and over a range of elevations, thereby sampling a large range in atmospheric optical depths.  With Bolocam, an AC biasing and demodulation technique is used to
read out the bolometers.  Fluctuations in the carrier amplitude
correspond to changes in the bolometer resistance in response to
changes in optical power.  On long timescales ($> 1$ scan),
variations in atmospheric opacity, or loading, cause changes in flux density calibration from both the signal attenuation and the change in the bolometers' responsivities induced by the loading changes.  Thus, the bolometer
resistances can be used to track calibration on a scan-by-scan basis.  The 
flux densities were calibrated from the loading for each observation using the 
composite calibration curve built up from the entire run.  This method is superior to photometric calibration
procedures that require interpolation in time or extrapolation 
across large ranges in elevation.

\subsection{Iterative Mapping and Sensitivity to Extended Sources}\label{iter}

The data were reduced using an IDL-based reduction package written for Bolocam (http://www.cso.caltech.edu/bolocam/).
The basic pipeline was described in \citet{lau05} and procedures 
unique to observations of molecular clouds (large area maps with high dynamic 
ranges and extended structures) were described in Paper I.  Here we summarize 
the techniques and refer the reader to those papers for details and 
results of simulations that characterized the algorithms.  

With Bolocam's AC bolometer bias and lock-in scheme, observations are made by scanning the telescope without 
chopping the CSO subreflector.  The AC bias and rapid scanning modulate the signals above the atmospheric $1/f$ knee; residual atmospheric emission (``background'') is removed by sky subtraction, or ``cleaning''. 
Bolocam was designed so that the bolometer beams overlap in the telescope 
near field, resulting in predominantly common-mode atmospheric background.  
This background was subtracted 
using a principle component analysis (PCA) that removed signals that were common mode across the focal plane array from the 
raw bolometer voltage time streams.  While sky background is effectively removed in this manner, source flux densities are 
attenuated too, especially in the case of extended sources which have 
significant common-mode components.  The first common-mode component is 
essentially the mean of all the bolometer signals.  Removal of successive 
orthogonal components subtracts more sky background, but each component  also
subtracts more from the source flux densities.  
Experimentation showed that removal of three PCA components removed the 
majority of the striping and sky background, 20\% to 30\% more than a simple mean 
subtraction.

Cleaning with bright sources in the field results in negative sidelobes 
along the scan direction because 
the mean timestreams, obtained by averaging over all bolometers, are set to zero by subtraction
of the first common-mode component.  Thus, the sources sit in 
negative bowls and the total flux densities for large (120\arcsec) bright
sources can be attenuated by as much as 50\%. 
This problem is remedied with an iterative mapping procedure in 
which sky subtraction is successively improved by removing a source model 
from the raw data and re-cleaning to recover structure missing from the 
original map (see Paper I). 

Simulations with fake sources of 
various sizes and signal-to-noise ratios showed that five or fewer 
iterations yield a nearly complete recovery of flux densities for 
bright point sources ($\sim98$\% recovery).  Flux densities of large, 
faint sources  with widths several times the beam size and amplitudes 
at approximately the level of the map rms noise are not recovered as well, 
with as much as 10\% of the flux density missing after 20 iterations.  
However, we restrict our photometry to small apertures and 
enforce a 4-$\sigma$ detection threshold, so the impact on our catalog is 
small.  We adopt a conservative 5\% photometric uncertainty from sky 
subtraction after iterative mapping.  All source flux densities will be 
biased low, but by less than this amount.

The iterative mapping procedure was run for each section of 
Ophiuchus and for each observing run. The May/June 2003 data and May 2004 
were iteratively mapped separately because they required different calibration 
and pointing corrections. The final map is a weighted average of
the maps from each run.

\subsection{Source Identification}\label{sourceid}

After the final calibrated maps were created, sources were identified as in Paper I.  
First, the maps were optimally filtered for point source detection, using a 
filter accounting for the beam size, telescope scan rate, and system noise 
Power Spectral Densities (PSDs), hereafter a Wiener filter, 
as described in Paper I.  The Wiener filtering attenuates the 1/$f$ 
noise and reduces the rms per pixel by about $\sqrt{3}$.  
Then, the map was 
trimmed to cut off the edges where the coverage was low. 
The coverage is dependent upon the number of observations, the 
number of bolometers that were read out, and the scan strategy.  
The average coverage in the map varies from 40 hits per pixel 
(a hit means that a detector passed over this area on the sky),
corresponding to 20 s of integration time, 
in most of the L1688 region to 60 (30 s of integration time) in the 
northeastern streamer.  If the coverage was less than 0.22 times the 
maximum coverage (corresponding to a range in the local rms over the 
map of a factor of $\sim2$) then that part of map was not included in the 
analysis.  Finally, a source-finding routine found all the peaks in the 
Wiener-filtered map above 4 $\sigma$ in the rms map.

A detection limit as low as 4 $\sigma$ was necessary to detect some previously
known sources, but this limit was not low enough to eliminate all 
false detections 
identified by the source extraction procedure.  Some artifacts were 
incorrectly identified as sources. Therefore, each source was 
inspected by eye.  Most of the artifacts were unambiguous because they were 
found close to the edges of the map or were caused by striping:  one 
pixel wide and extended in one of the scan directions. Single-pixel peaks 
were also discarded, which might have resulted in excluding some 
faint sources.
Although in principle it should be possible to recover structure up to the 
array size of $7\farcm5$, it was found in Paper~I that 
structures $\gtrsim 4\arcmin$ are severely affected by cleaning, and not 
well recovered by iterative mapping.

\section{Results}

\subsection{General cloud morphology}

\begin{figure}
\plotone{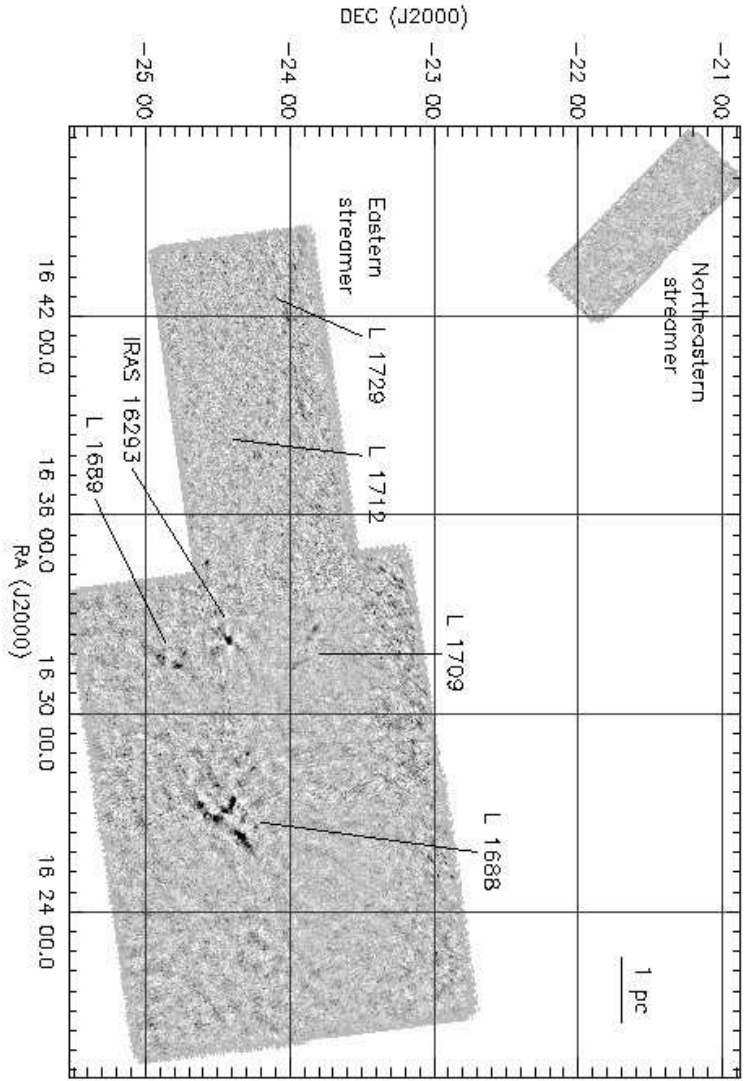} \figcaption{\label{ophall} 1.1 mm Bolocam map
of 10.8 deg$^2$ (51.4 pc$^2$ at $d=125$~pc) in the Ophiuchus molecular 
cloud, with $10\arcsec$ pixels and a beam size of $31\arcsec$.  
The gray scale is proportional to intensity weighted by the coverage to
avoid confusion by noise in regions with low coverage.
Well known regions and those discussed by name in the text are indicated.  
}
\end{figure}

The map of the cloud is shown in Figure~\ref{ophall}, with known regions 
identified. 
Our map covers 10.8 deg$^2$ (51.4 pc$^2$ at a distance of 125~pc), 
which is equivalent to $1.4\times10^5$ resolution elements given the 
beam size of $31\arcsec$.
Most of the compact emission is confined to the L1688 cluster region.  
Several sources are also detected in L1709, L1689, and around the 
extensively studied Class~0 protostar \irassrc.  
No emission that is extended $\gtrsim 2\arcmin$ is seen in the map.  

\begin{figure}
\plotone{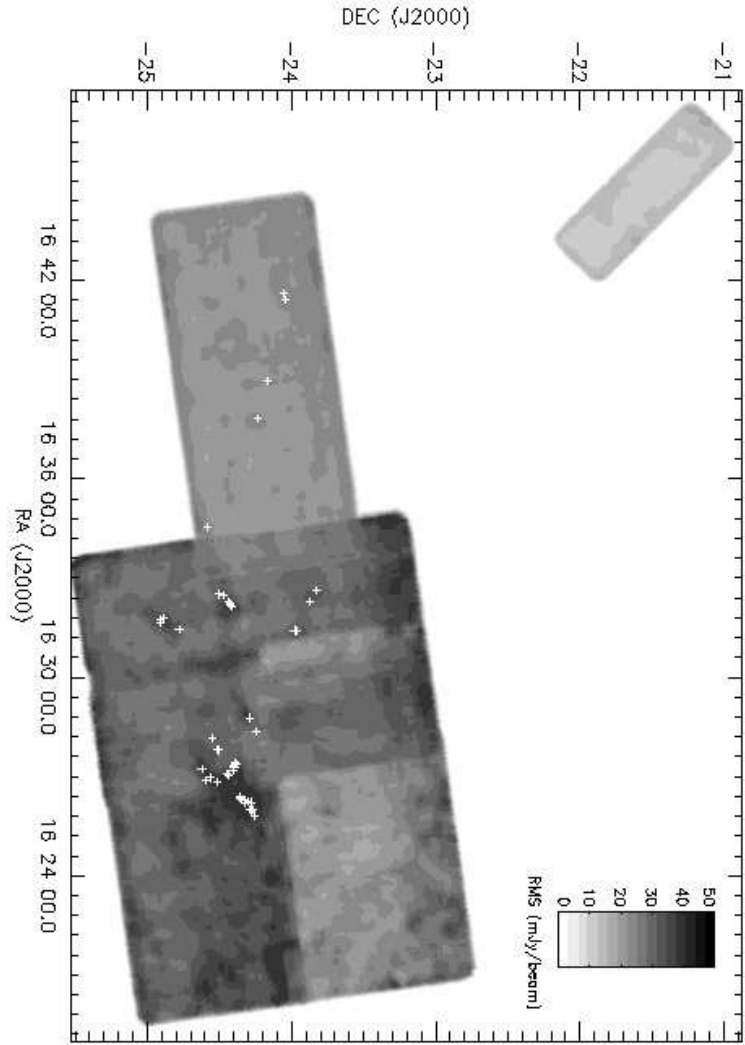} \figcaption{\label{noisefig} 
A map of the noise in gray scale with sources indicated by white plus signs.
The gray scale runs from 11 mJy beam$^{-1}$ to 30 mJy beam$^{-1}$.
High noise regions are apparent in a strip above L1709, and in the 
area containing L1688.
Note the increased noise near bright sources caused by residual
systematics from sky subtraction.
}
\end{figure}

The noise in the final map varied from section to section
because of differences in the number of good observations and changes in sky 
noise.  A map of the noise (Fig. \ref{noisefig})
shows the variations in noise in the different map areas, ranging from
11 to 30 mJy beam$^{-1}$. The average rms in the regions
of the map where most sources were detected was about 27 mJy beam$^{-1}$. 
High noise regions are apparent in Figure~\ref{noisefig} 
as a strip above L1709 and in the regions around strong sources, 
especially in the L1688 cluster.

\begin{figure}
\plotone{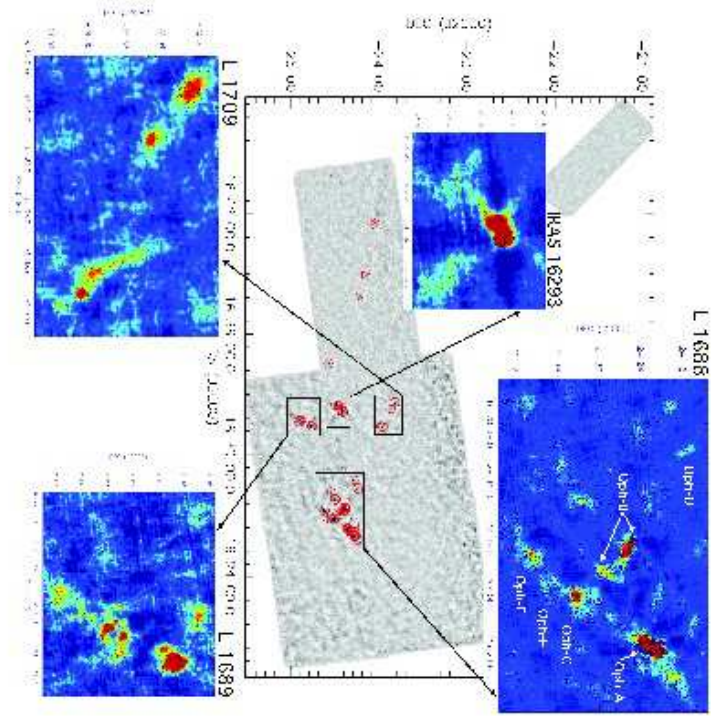} \figcaption{\label{ophsource} 1.1 mm
Bolocam map of the Ophiuchus molecular cloud, with the positions of the 44 
sources detected above 4 $\sigma$ marked as circles.  
The gray scale shows the intensity {\it not} weighted by the coverage.
The inset maps show particular regions on an expanded scale. The
conversion from intensity to color differs among the insets to cover
the large range of intensity.
Sources marked by triangles in the eastern streamer are below the 4-$\sigma$
detection limit so are tentative detections (but see Fig. \ref{sst3}).
}
\end{figure}

We detected 44 sources with signal-to-noise greater than 4 $\sigma$ that were
confirmed as real by inspection. These are listed in Table \ref{sourcetab} and
identified for convenience as Bolo~1, etc. 
All of these sources were identified in the
main cloud and eastern streamer sections. Figure
\ref{ophsource}  plots the positions of the  sources as
red circles on the grayscale 1.1~mm map, with insets showing 
magnifications of the densest source regions. We did not detect any
sources in the northeastern streamer, where the noise is lowest.  
Most of the sources are concentrated in the previously
well-studied regions in Ophiuchus, 
suggesting that dense cores are highly
clustered in the Ophiuchus cloud. Figure \ref{ophsource} contains
blow-ups of the main regions of emission, including the well known
Ophiuchus cluster in L1688, L1689, \irassrc, and L1709. 

Visual comparison to 
previous maps of dust continuum emission in the L1688 cluster 
indicates reasonable agreement on the overall shape of the emission, 
considering differences in resolution \citep{mot98} and 
wavelength \citep{joh00}. However, detailed comparison of source positions
in Table \ref{sourcetab} and those in \citet{joh00} shows that a substantial
number of our sources are separated into multiple sources by \citet{joh00},
who used the the clumpfind algorithm on data with better resolution by
a factor of two; in the same area, they list 48 sources 
compared to our 23. The list of 48 includes some small, weak, but unconfused
sources that we do not see. Assuming that $S_\nu \propto \nu^3$, 
as expected for emission in the Rayleigh-Jeans limit with an opacity
proportional to $\nu$, some of these sources
should still lie above our detection limit, but not far above.

The \citet{stanke05} map covers 1.3 deg$^2$ of L1688 (slightly less
than one of the boxes defined by the grid lines in Fig. 2), with
lower noise ($\sim 10$ mJy) and a slightly smaller beam (24\arcsec )
at nearly the same wavelength (1.2 mm). Their images are qualitatively
very similar to the L1688 inset image in Fig. 4. However, they find
143 sources in this region, using wavelet analysis and clumpfind,
and by essentially cleaning down to the noise. They include sources that
are less than 3 $\sigma$ but extended. These differences make it
difficult to compare sources in detail, but it appears that many of
our sources would be split into multiple sources by \citet{stanke05}.

Another useful comparison is with the work of \citet{visser02} in a less
crowded area of Ophiuchus. They found 5 sources along a filament in L1709;
we find three sources in reasonable agreement in position, while L1709-SMM3 and
L1709-SMM5 from their paper are blended into Bolo~30 in Table \ref{sourcetab}.
We see additional structure below the 4-$\sigma$ limit extending to the
northeast of that group of sources that is not seen in the \citet{visser02}
map.
The most diffuse source in their map, L1709-SMM4, shows up strongly in our
map, but shifted about 25\arcsec\ east, an example of position shifts caused
by different sensitivity to large scale structure and source finding 
algorithms. \citet{visser02} also found a weak source in L1704 that we do
not see, consistent with our detection limit. These points should be
borne in mind when we compare source statistics to those of previous
work in later sections.

Several diffuse emission peaks were observed in the eastern
streamer, an area that includes L1712 (16\h\ 38\m, $-$24\degree\ 26\arcmin) and
L1729 (16\h\ 43\m, $-$24\degree\ 06\arcmin). However, these cores,
though visible by eye in the map, are only 3-$\sigma$
detections and are listed separately as tentative detections in Table 1 and are 
not included in our source statistics.  These sources are
in a long filament of extinction that extends east from the main cloud
(\citealt{cam99}; \citealt{ridge06}). 
We believe at least some of these sources are real, based on inspection by 
eye and comparison to \spitzer\ maps of the region.  
In Figure \ref{sst3}, the tentative 1.1~mm sources align with an elongated 
structure that is  dark at 8 \micron, but bright at 160 \micron, 
suggestive of a cold, dense filament.
This filament was previously observed in $^{13}$CO
\citep{lor89} and C$^{18}$O \citep{tac00}, but it  has not been mapped in the
millimeter continuum until now. While the overall morphology is similar to that
seen in C$^{18}$O \citep{tac00}, only Bolo~45 has an obvious counterpart,
$\rho$-Oph~10, in the table of \citet{tac00}.

The most striking feature of the Bolocam map of Ophiuchus is the lack
of 1.1~mm emission in regions outside of known regions of star formation, 
even in areas with significant extinction ($A_V >$ 3 mag). 
Figure \ref{avmap} shows the Bolocam map of Ophiuchus overlaid with extinction
contours constructed using the {\it{NICE}} method (e.g., \citealt{lada99};
\citealt{huard06}), making use of 2MASS sources, and convolving the 
line-of-sight extinctions with a Gaussian beam with FWHM of 5\arcmin.  
This method depends on background stars to probe the column densities 
through the cloud.  Similar to Enoch et al. (2006), we eliminate from the 2MASS
catalog most foreground and embedded sources that would yield
unreliable extinction estimates when constructing the extinction map.  
In order to calibrate the extinction map, we identified two 
``off-cloud'' regions, which were free of structure and assumed to 
be non-extincted regions near the Ophiuchus cloud.  These off-cloud regions 
contained a total of more than 13000 stars and were 
0$\ptdeg$6 $\times$ 0$\ptdeg$6 and 1$\ptdeg$5 $\times$ 0$\ptdeg$2
fields centered on $\alpha $= 16$^h$44$^m$00$^s$, 
$\delta $= $-$22$^\circ$54$^\prime$00$^{\prime\prime}$ 
and $\alpha $= 16$^h$39$^m$12$^s$, 
$\delta $=$-$25$^\circ$24$^\prime$00$^{\prime\prime}$ (J2000.0), respectively.
The mean intrinsic H$-$K color of the stars in these off-cloud fields
was found to be 0.190 $\pm$ 0.003 magnitudes. We assume $\av = 15.9$
E(H--K) to convert to \av\ \citep{rieke85}.

The 1.1 mm sources are all in regions of high extinction, but not all
regions of substantial extinction have Bolocam sources.  
For example, we found no 1.1 mm sources in the small northeastern streamer of
Ophiuchus that could be confirmed as real by eye despite having much
lower noise in this region than for the rest of the map.  
The beam-averaged extinctions in the
northeastern streamer are $A_V \approx 3$ to 8 magnitudes.
The 4-$\sigma$ detection limit in this region 
corresponds to objects with masses as small as 0.06 \msun\ (see
\S \ref{stats}).
Thus even in relatively high extinction regions, much of the Ophiuchus cloud 
appears devoid of dense cores down to a very low mass limit.

\subsection{Source Properties}\label{stats}

\subsubsection{Positions and Photometry}\label{photsec}

Table~\ref{sourcetab} lists the position, peak flux density, and
signal-to-noise ratio (S/N) of the 44 4-$\sigma$ sources, and the 
four 3-$\sigma$ detections in the eastern streamer are listed separately.  
All statistical analysis is based on the 44 secure detections only.
For known sources the most common names from the literature are also given.  
Some are known to host protostars while others may be starless.
The peak flux density is the peak pixel value in the $10\arcsec$ pixel$^{-1}$ 
unfiltered map (without the Wiener filter applied).
The uncertainty in the peak flux density is the local (calculated
within a 400\arcsec\ box) rms beam$^{-1}$
and does not include an additional 15\% systematic 
uncertainty from calibration uncertainties and residual errors after 
iterative mapping.
The S/N is calculated from the peak in the Wiener-convolved map compared 
to the local rms because this is the S/N that determines detection.

Source photometry is presented in Table~\ref{phottab}.
Aperture photometry was calculated using the IDL routine APER.  
Flux densities for each source are given in Table~\ref{phottab}
within set apertures of 40\arcsec, 80\arcsec, and 120\arcsec.  
If an aperture is larger than the distance to the nearest neighboring source, 
the flux is not given, to avoid contamination of fluxes by nearby sources.
Table~\ref{phottab} also lists the total flux density, which is integrated 
over a 160\arcsec\ aperture or the largest aperture up to 160\arcsec\
that does not include flux by nearby sources (defined such that the aperture 
radius is less than half the distance to the nearest neighbor).
The uncertainties for the flux densities are
$\sigma_{ap} = \sigma_{beam} \sqrt{\frac{A_{ap}}{A_{beam}}}$, where
$\sigma_{beam}$ is the local rms beam$^{-1}$, and $A_{ap}$ and 
$A_{beam} = \pi \beam^2/(4 \ln 2)$ are the aperture and beam areas. 
The uncertainties in Table~\ref{phottab}
do not include an additional 15\% systematic uncertainty in the flux
densities that results from the absolute calibration uncertainty (10\%)
and systematic biases remaining after iterative mapping (5\% for
bright sources). 

The distribution of flux densities for the 44 detected sources is shown in 
Figure~\ref{flux}. This figure compares the distribution of peak flux 
densities to the total flux densities.  The peak flux density  distribution 
has a mean of $\sim0.6$~Jy beam$^{-1}$.  The total flux density
distribution has a mean of about 1.6~Jy.  The shaded region in 
Figure~\ref{flux} indicates the 4-$\sigma$ detection limit, which 
varies throughout the map from $\sim 0.06-0.12$~Jy beam$^{-1}$.  
The flux density distributions shown in Figure \ref{flux} are 
similar to the distributions of peak and total flux densities of 
Bolocam sources in Perseus (Paper I) in that the total flux 
density distribution is shifted from the peak distribution 
because most sources are larger than the beam.

\subsubsection{Sizes and Shapes}\label{sizes}

Sizes listed in Table~\ref{phottab}
 were found by fitting a 2D elliptical Gaussian to 
determine the FWHM of the major and minor axes and the position angle
of the ellipse (PA), measured east of north. 
The errors are the formal fitting errors. There are additional
uncertainties due to noise and residual cleaning effects on the order
of 10 -- 15\% in the FWHM and $\sim5$\degree\ in the PA. The size of the
source is limited by the distance to its nearest neighbor, because all
emission at radii greater than half the distance to the nearest
source is masked out for the Gaussian fit to avoid including flux from 
neighboring sources in the fit. This procedure also ensures that the size and 
the total flux density of a source are measured in approximately the same
aperture. The sources in the L1688 cluster are quite crowded and source
sizes and fluxes may be affected by nearby sources.

Figure \ref{size} shows the distributions of the major and
minor axis FWHMs from the fits.  Both distributions peak
between 50\arcsec\ and 60\arcsec, and the average axis ratio is
1.2. Only a few sources have large FWHM sizes ($>100\arcsec$).  
Many of the sources in the map are part of filamentary
structures. Individual sources are found by identifying peak pixels
above a 4-$\sigma$ cutoff. Therefore, large clumps of emission are
often broken down into several smaller sources, because the clumps
contain several peaks.  This method of finding cores and the
filamentary nature of the dense gas in Ophiuchus could result in a
slight elongation of the sources. However, the majority (61\%) of
sources in the entire sample are not elongated at the half maximum 
level (axis ratio $<$ 1.2), 
suggesting roughly spherical condensations along the filaments.

Morphology keywords for each source are also listed in Table
\ref{phottab}, which indicate if the source is multiple (within
3\arcmin\ of another source), extended (FW at 2 $\sigma >$ 1\arcmin),
elongated (axis ratio at 4 $\sigma >$ 1.2), or weak (peak flux density is less
than 8.7 times the rms beam$^{-1}$.
Of the 44 sources, 18 are classified as round, with axis ratio at the
4-$\sigma$ level less than 1.2. The difference between this result and
the fact that 61\% had axis ratios from the Gaussian fits $< 1.2$ indicates
relatively round sources within more extended, elongated structures;
this result suggests spherical sources embedded along filaments.
Visual inspection indicates that the elongated lower contours are usually
elongated along the local filametary structure, as seen both in the still
lower contours and in the extinction map (Fig. \ref{avmap}).
Future polarimetric observations could determine the role of magnetic fields
in the filamentary structures.
Of the 44 sources, 36 are multiple, reflecting the strongly clustered
nature of the sources (see later section). Also 36 sources are extended.
Only two sources are neither multiple nor extended. We see no evidence for
a population of isolated, small, dense cores.

\subsubsection{Masses, Densities, and Extinctions}\label{masssec}

Isothermal masses for the sources were calculated according to the 
equation
\begin{equation}
M = \frac{D^2 S_{\nu}}{B_{\nu}(T_D) \kappa_{\nu}} \label{masseq}
\end{equation}
where $D$ is the distance (125 pc), $S_\nu$ is the total flux density, $B_\nu$
is the Planck function, $T_D$ is the dust temperature, and
$\kappa_\nu$ is the dust opacity.
We interpolate from the dust model of
\citet{oss94} for grains with coagulated ice mantles (their Table 1,
column 5), hereafter referred to as OH5 dust, to obtain
$\kappa_\nu$ = 0.0114 cm$^2$ g$^{-1}$ at 1.12 mm. 
This dust opacity assumes a gas to dust mass ratio of 100,
so the above equation yields the total mass of the molecular
core. Table \ref{phottab} lists the isothermal masses for a dust
temperature of 10 K. The uncertainties in the masses listed in Table
\ref{phottab} are from the uncertainty in the total flux density.
The total uncertainties include uncertainties in distance,
opacity, and \td, and are at least a factor of 4 
\citep[][and see Paper~I for a more complete discussion]{shir02,young03}.

The total mass of the 4-$\sigma$ sources is 42 \msun, 
with $\mean{M} = 0.96$ \msun, and a range from $0.24$ to $3.9$ \msun.
The \citet{joh04} survey of Ophiuchus found a larger total mass (50 \msun) 
in a smaller area than we covered. Part of the difference results from
their assumption of a larger distance (160 pc), which would make all their
masses larger than ours by a factor of 1.6. However, they assume a value for
$\kappa_{\nu}$ at 850 \micron\ that is 1.1 times higher than OH5 dust,
and they assume $\td = 15$ K, which would decrease our masses by a factor
of 1.94. If we used the assumptions of \citet{joh04} for our data,
we would derive a total mass of 32 \msun\ from our data.
The source of the difference between our result for total mass in cores
and that of \citet{joh04} does not seem to be explained by the assumptions 
used to obtain mass. More likely, it arises from differences in methods
of defining sources. If we integrate all the areas of the map above
4 $\sigma$, we get 131 Jy, which translates to 79 \msun, using our
usual assumptions, or 60 \msun, using the assumptions of \citet{joh04}.
Thus about half the mass traced by emission cannot be assigned to a
particular core, mostly because it is in confused regions.
\citet{deg89} estimate a total mass in Ophiuchus of $10^4$ \msun\, 
while we find a total of 2300 \msun\ above $\av = 2$ (Table \ref{tab3}).
The percentage of cloud mass in dense cores is between  0.4\% and
1.8\%. This fraction is even lower
than that found in in Paper~I for Perseus (between 1\% and 3\%).

The mean particle density for each source is estimated 
as $\mean{n} = M/((4/3) \pi R^3 \mu_p m_H)$, where $M$ is the total mass, 
$R$ is the mean deconvolved HWHM size, 
and $\mu_p=2.37$ is the mean molecular weight per particle, including
helium and heavier elements.  
The mean densities are quite high compared to the surrounding
cloud, ranging 
from $\mean{n} = 9\ee4$ \cmv\ to 3\ee7 \cmv, with an average value of 
$1.6\ee6$ \cmv. The free-fall timescale for the mean density would be
only $2.7\ee4$ yr.

The beam-averaged column density of H$_2$ at the peak of the emission
is calculated from the peak 1.1~mm flux density $S_{\nu}^{beam}$:
\begin{equation}
N(\mathrm{H}_2) = \frac{S_{\nu}^{beam}}{\Omega_{beam} \mu_{H_2} m_H \kappa_{\nu} B_{\nu}(T_D)}. \label{aveq}
\end{equation}
Here $\Omega_{beam}$ is the beam solid angle, and  $\mu_{H_2}=2.8$ is the 
mean molecular weight per H$_2$ molecule, which is the relevant quantity for
conversion to extinction.
We assume a conversion from column density to 
$A_V$ of $N($H$_2)/A_V =  0.94 \times 10^{21}$~cm$^{2}$~mag$^{-1}$ 
\citep{bohlin78}, using $R_V = 3.1$. This relation was determined in the diffuse
interstellar medium and it may not be correct for such highly
extinguished lines of sight as we are probing.
Peak extinctions range from 11 to 214 mag for the 4-$\sigma$ sources with
a mean value of 43, while
the tentative detections in the eastern streamer range from 7 to 11.


The extinctions within the cores should be distinguished from the
surrounding extinction, as traced by the {\it{NICE}} method with 2MASS 
sources.  While 2MASS sources probe the low-to-moderate extinctions 
within the Ophiuchus cloud, the sensitivity of the 2MASS observations 
is not sufficient to probe reliably the high extinction regions 
traced by the millimeter emission.  By considering both tracers of 
extinction, the morphology of the cloud can be inferred over a 
large range of column densities, from the diffuse and vast regions 
of the cloud (containing most of the mass) to the densest cores.
Almost all the 1.1 mm emission lies within the contour of $A_V = 10-15$ mag,
as determined from the near-infrared (Fig. \ref{avmap}); the faint 
emission in the eastern streamer is the main exception. A quantitative
comparison will be made in \S \ref{threshold}.

\subsection{Comparison to Perseus (Paper~I) \label{compsect}}

One considerable advantage to conducting a survey of several regions with the 
same instrument is the elimination of a number of biases that can result 
from observations with different instruments or at different wavelengths.  
Ophiuchus is the second cloud in a series of three large 1.1~mm surveys using 
Bolocam, including Perseus (Paper~I) and Serpens (Enoch et al., in prep).  
Here we compare the results for Perseus and Ophiuchus.

The distribution of source sizes is considerably different for Ophiuchus and 
Perseus, especially if one considers the deconvolved source sizes, as 
illustrated in Figure~\ref{compsize}, which plots the fractional number of 
total sources as a function of size.   Perseus sources have a larger mean 
linear size than those in Ophiuchus ($1.5\times 10^4$~AU 
vs $7.6\times10^3$~AU), and the Perseus distribution extends 
to $3\times10^4$~AU, twice the maximum size of sources in 
Ophiuchus ($1.5\times10^4$~AU). The fact that Ophiuchus is closer
(125 versus 250 pc) could play a role, but the distributions in both
clouds lie well above the resolution limits (vertical lines in 
Fig. \ref{compsize}). The largest recoverable size is about 240\arcsec,
which corresponds to 3\ee4 AU in Ophiuchus and 6\ee4 AU in Perseus, larger
than the relevant distributions in Fig. \ref{compsize}. However, in 
clustered regions, the size is limited by the nearest neighbors.
The fraction of sources classified as multiple is somewhat higher
(0.82) in Ophiuchus than in Perseus (0.73) and the ratio of mean separation
to mean size is smaller in Ophiuchus than in Perseus (2.5 versus 3.0).

A comparison of the distribution of axis ratios for the two 
clouds (Figure~\ref{compaxis}) shows that sources in Perseus tend to 
be more elongated than those in Ophiuchus. In particular, 
the Perseus distribution has a tail that extends to much greater axis 
ratios.  The mean axis ratio in Perseus is 1.4 compared to 1.2 in Oph.  
It was found in Paper~I that axis ratios smaller than 1.2 are not 
significantly different from unity; thus sources in Ophiuchus are 
round on average, while sources in Perseus are elongated on average. 
The larger, more elongated sources in Perseus may be a real effect, or 
may be due, at least in part, to the different distances to the clouds.
As noted above, many sources in Ophiuchus are round as measured by the
axis ratio, but elongated as measured by the lower contours, possibly
reflecting the effects of being in a filament. At greater distance, these small
round cores may not stand out above the elongated lower level emission.

The mean mass of 0.96 \msun\ is about half that in Perseus (2.3 \msun),
but the smaller size of the sources in Ophiuchus makes the mean density
higher ($\mean{n} = 1.6\ee6$ \cmv\ compared to $4.3\ee5$ \cmv\ in
Perseus) and the mean of the peak extinctions in the cores higher as well
($\mean{A_V} = 43$ mag versus 25 mag for Perseus).

Further comparison to Perseus will be incorporated into various
parts of \S \ref{discussion}.

\section{Discussion}\label{discussion}

In the following sections, we discuss issues of completeness in the context
of the mass-size relations, and then discuss the mass function of cores.
We then discuss clustering tendencies and the extinction threshold.

\subsection{Completeness}\label{completeness}

Figure \ref{mass_size} shows the distribution of source mass versus
size, where the size is the geometric average of the major and minor FWHM for
each source. The minimum detectable mass and source size are related because we
detect sources from their peak flux density, but calculate the mass from the
total flux density.  Therefore, we are biased against large, faint,
low-mass sources. For Gaussian sources, the mass calculated from the
total flux density is related simply to the mass from the peak flux density:
\begin{equation}
M_{limg} =  M_{limp} (\theta_s/\theta_b)^2 \left[1 - \exp(-4 \ln 2 
(120/\theta_s)^2 \right], 
\end{equation}
where $M_{limg}$ is the mass limit for a Gaussian source,
$M_{limp}$ is the mass limit for a point source, and
$\theta_s$ and $\theta_b$ are the FWHM of source and beam, respectively.
The last factor corrects for flux from sources larger than our largest
aperture, but has very little effect except for the few largest sources.
Using this relation, we compute a 50\% completeness
level for mass that varies with size (this is essentially a line with
$M \propto R^2$, where $R$ is the radius). 
This limit is indicated by the solid lines (showing the range of rms)
in Figure \ref{mass_size} (middle panel). 

The real mass completeness limit is more complicated, even for Gaussian 
sources, as a result of the reduction and detection processes applied to the 
data.  Empirical 10\%, 50\% and 90\% completeness limits are also 
plotted in Figure \ref{mass_size} (bottom, middle, and top panels, 
respectively). These limits were determined by introducing
simulated sources of various peak flux densities with FWHMs of 30\arcsec,
60\arcsec, 100\arcsec, and 150\arcsec\ into a portion of the
Ophiuchus data with no real sources. The data with the simulated
sources were then cleaned and iteratively mapped, and the same method
was used to extract the sources from the maps as for the real
data. The completeness limits indicate what percentage of simulated
sources with a particular size and mass were detected above 4 $\sigma$
in the Wiener-convolved map.
Because the local rms varies substantially across the map, completeness 
limits have been calculated both in low rms (20 mJy beam$^{-1}$, 
lower line in each panel) and high rms ( 25 to 30 mJy beam$^{-1}$, 
upper line) regions.  

Most of the 44 sources are found in the higher rms regions of the map,
corresponding to the upper curves in Figure \ref{mass_size}.
Some of this noise is caused by sidelobes, etc. of the very strong
sources (Fig. \ref{noisefig}).
Very large sources (FWHM $>100\arcsec$) are not fully recovered by the 
iterative mapping routine (see Paper~I), and therefore tend to 
have a higher mass limit than expected for a simple scaling with source 
size.  This is illustrated in the middle panel of Figure~\ref{mass_size}, 
where the empirical completeness limit (dash-dot line) rises above the 
Gaussian limit (solid line) for large sources.
Typical 1-$\sigma$ error bars in $M$ and FWHM are shown for $50\arcsec$ 
and $100\arcsec$ FWHM sources near the detection limit.  
The uncertainty in mass is from the uncertainty in the integrated 
flux (including the 5\% uncertainty from the cleaning process, but not 
the absolute calibration uncertainty), and $\sigma_{\mathrm{FWHM}}$ 
is estimated from simulations.

The mass-size relation does not look like a distribution of constant
density cores of varying sizes ($M \propto R^3$) nor such a collection
of cores with constant column density ($M \propto R^2$). Rather, it looks
as if there are two populations, with different sizes but, given the
completeness limitations, similar masses. The compact sources have a wide
range of masses and mean densities.

\subsection{The Core Mass Distribution}

Figure \ref{cum_mass} shows the differential ($dN/dM$) core mass distribution 
(CMD) for the 44 secure detections.
These include both starless cores and cores with protostars.
The masses are taken from Table \ref{phottab}.
Error bars in Figure \ref{cum_mass} are $\sqrt{N}$ statistical errors only.
The shaded regions on the figure represent the range in detection limit for a 
point source (left), and the 50\% completeness limit for sources with a 
FWHM of $\sim 70\arcsec$ (right), which is approximately the average 
FWHM of the sample.  We do not attempt to correct for incompleteness 
in the mass function.  Most sources are found in the higher noise regions 
of the map; therefore the mass function is likely to be incomplete
below 0.5 \msun.  

The CMD above 0.5 \msun\ can be fitted with either
a power law \citep{sal55}, which gives a reduced chi-squared of $\chisq = 0.4$, 
or a lognormal function \citep{miller79}, which gives $\chisq = 0.3$.
The slightly better \chisq\ value for the lognormal function reflects the
tendency of the distribution to flatten at lower masses, but incompleteness 
prevents us from distinguishing between these two functions.
For the power law ($N(M) \propto M^{-\alpha}$), the best fit is for
$\alpha = 2.1\pm 0.3$.
The lognormal distribution is given by
\begin{equation}
\frac{dN}{dlogM} =  A exp \left[ \frac{-(logM-logM_0)^2}{2\sigma_M^2} \right], 
\end{equation}
where $A$ is the normalization, $\sigma_M$ is the width of the distribution, 
and $M_0$ is the characteristic mass.  The best-fitting lognormal function
for $M>0.5$ \msun\ has $\sigma_M=0.5 \pm 0.4$ and $M_0=0.3 \pm 0.7$ \msun.

The CMD depends on assumptions about distance, opacity,
and dust temperature.
Increasing the distance shifts it to higher masses, while 
increasing the opacity or dust temperature shifts it to lower masses.
The CMDs in Perseus for four different dust temperatures ($T_D= 5$, 10, 20, 
and 30 K) are shown in Figure 17 of Paper I.
The distribution moves to lower masses for increasing
temperature, but the overall shape of the distribution is not affected
by changing \td. However, if \td\ varies systematically with mass,
the shape of the distribution could be changed. Experiments in which
cores in the main cluster were given higher temperatures, or small cores
were given higher temperatures produced little change in the mass
distribution. If cores in the L1688 cluster were assigned $\td = 20$ K
and other cores assigned $\td = 10$ K, the best-fit value became
$\alpha = 2.2$ for $M>0.5$ \msun, insignificantly different. However,
the evidence for a turnover at low masses became even less
significant. Such effects should be considered before inferring 
turnovers in CMDs.

\citet{joh00} (see their figure 7)
fit the cumulative mass distribution for 850 \micron\ cores within the
L1688 region, assuming $\td = 20$ K, with a broken power law. They found
$\alpha_1$ = 1.5 for masses less than
about 0.6 \msun\ and $\alpha_2$ = 2.5 for M~$>0.6$~\msun. 
The \citet{joh00} sample is complete down to about $M \sim$ 0.4 \msun. 
If we assume $\td = 20$ K, the  best-fit power law slope remains
$\alpha = 2.1$, but our completeness limit becomes 0.2 \msun.
Thus, our mass function declines less rapidly than that of \citet{joh00},
but the difference is not very significant.  Since \citet{joh00} split
some of our sources into multiple, smaller sources, it is natural that they
would find a larger value of $\alpha$.
\citet{stanke05} do not give a table of masses, but their CMD extends up
to roughly 3 \msun, similar to our result, despite differences in source
identification and mass calculation.
They argue for breaks in their CMD around 0.2 and 0.7 \msun, with
$\alpha \sim 2.6$ for large masses.

The CMDs for Ophiuchus and Perseus are 
shown together for comparison in Figure~\ref{compmass}, where the 
Perseus distribution has been scaled down by a factor of five to 
match the amplitude of the Ophiuchus mass function.  In the region 
where both mass functions are reasonably complete ($M>0.8$ \msun), the 
two distributions appear quite similar except for the fact that the 
Ophiuchus mass function drops off at $\sim 4$ \msun\ whereas the 
Perseus mass function extends to $M >10$ \msun. With a factor of 3 more
sources in Perseus, statistical sampling of the same mass function naturally
results in a higher maximum mass.
The best single power-law fit ($\chisq = 0.8$) to the Perseus distribution 
for $M>0.8$ \msun\ gave $\alpha=2.1\pm 0.15$, the 
same as the slope for Ophiuchus for $M>0.5$ \msun.
\citet{TS98} also found $\alpha = 2.1$ for a cluster in Serpens.
\citet{mot98} found $\alpha = 2.5$ above 0.5 \msun\ for a broken power-law 
fit to cores in the Ophiuchus cluster, and \citet{joh00} found a similar
result. 
Broken power-law fits tend to produce steeper
slopes at higher masses, and the slopes are steeper if a higher break
mass is assumed, suggesting that lognormal fits may be appropriate.
The best lognormal fits to the Ophiuchus ($\sigma_M = 0.5 \pm 0.4$, 
$M_0=0.3 \pm 0.7$ \msun ) and Perseus ($\sigma_M = 0.5 \pm 0.1$, 
$M_0=0.9 \pm 0.3$ \msun ) mass functions 
have similar shapes within the uncertainties.

The CMD is most naturally compared to predictions
from models of turbulent fragmentation in molecular clouds. 
\citet{padoan02} argue that turbulent fragmentation naturally 
produces a power law with $\alpha = 2.3$ (for the differential CMD that
we plot). However, \citet{paredes06} question this result, showing that
the shape of the CMD depends strongly on Mach number in the turbulence.
As the numerical simulations develop further, the observed CMD will provide
a powerful observational constraint, with appropriate care in turning the
simulations into observables.

The shape of the CMD may also  be related to the process 
that determines final stellar masses.   
Assuming the simplest case in which a single process dominates the 
shape of the  stellar initial mass function (IMF), the IMF should 
closely resemble the original CMD if stellar masses 
are determined by the initial fragmentation into cores \citep{AF96}.  
Alternatively, if stellar masses are determined by other processes,
such as further fragmentation within cores, merging of cores,
competitive accretion, or feedback, the IMF need not be related
simply to the CMD (e.g., \citealt{paredes06}).

In addition, the IMF itself is still uncertain \citep{scalo06}.
For example, the Salpeter IMF would have $\alpha = 2.35$ \citep{sal55} in
our plots.
More recent work on the local IMF finds evidence for a break in the
slope around 1 \msun. The slope above the break depends on the choice
of break mass. For example, \citet{reid02} find $\alpha = 2.5$ above
0.6 \msun, and $\alpha = 2.8$ above 1 \msun. \citet{chab03} suggests
$\alpha = 2.7$ ($M>1$ \msun), while \citet{schroder03} finds 
$\alpha = 2.7$ for $1.1 < M_\star < 1.6$ \msun and $\alpha = 3.1$ for
$1.6 < M_\star < 4$ \msun.
Given the uncertainties and the differences
between fitting single and broken power laws, all these values for
$\alpha$ are probably consistent with each other and with determinations of
the CMD.

Currently, we cannot separate prestellar cores from more evolved objects in 
either Perseus or Ophiuchus, so a direct connection to the IMF is 
difficult to make. After combining these data with \spitzer\ data 
it will be possible to determine the evolutionary state of each source 
and compare the mass function of prestellar cores only.
Further comparisons of clustering properties and the probability of 
detecting cores as a function of $A_V$ in Ophiuchus and Perseus 
are discussed in the following sections.

\subsection{Clustering}

The majority of the sources detected with Bolocam in Ophiuchus are
very clustered.  Of the 44 sources, 36 are multiple (Table \ref{phottab}),
with a neighboring source within 3\arcmin, corresponding to 22,500~AU
at a distance of 125 pc. The average separation for the
whole sample is 153\arcsec, or 19000~AU.  
If  we consider only sources in the L1688 region for comparison to 
previous studies, the mean separation is $116\arcsec$, or 14500~AU. 
The median separation in L1688 is substantially smaller 
($69\arcsec = 8600$~AU).
The median separation in L1688 is very similar to the mean size of the
sources in the sample, 68\arcsec, as determined by averaging the major
and minor FWHM.  This indicates that many source pairs are barely resolved.  
It also means that the measured size of many sources is limited to something 
like the mean separation, since the Gaussian fitting routine takes into account
the distance to the nearest neighbor when determining source size. 

The median separation of 8600 AU for the L1688 cluster is only slightly 
larger than the fragmentation scale of 6000 AU  suggested
by \citet{mot98} in their study of the main Ophiuchus cluster by
examining the mean separation between cores in their data.  
Resolution effects likely play a role here, as our resolution (3900 AU)
is approximately twice that of \citet{mot98}.
\citet{stanke05} find two peaks in the distribution of source separations of
neighboring cores ($\sim 5000$ AU and $\sim 13000$ AU), suggesting that
they also distinguish the cores in the Ophiuchus cluster from those in the
more extended cloud.
The median core separation is still smaller than the median separation of T
Tauri stars in Taurus of 50000 AU \citep{gom93} as pointed out by
\citet{mot98}.

Another description of source clustering is provided by
the two-point correlation function,  as was used in Paper I and
\citet{joh00}. Figure \ref{hr} plots $H(r)$, $w(r)$, and log($w(r)$)
versus the log of the distance between sources, $r$.  $H(r)$ is the
fractional number of source pairs with a separation between 
log($r$) and log($r$) + dlog($r$) 
and is plotted both for the Ophiuchus sources ($H_s(r)$;
solid lines in Figure \ref{hr}) and for a uniform random distribution of sources
($H_r(r)$; dashed lines) with the same observational RA and Dec limits as
the real sample (i.e. there are no sources in the random sample 
outside the actual area observed).   
Because it is discontinuous from the rest of the map, the northeastern 
streamer is not included in this analysis.
$w(r)$ is the two-point correlation function, given by
the equation:
\begin{equation}
w(r)=\frac{H_s(r)}{H_r(r)} - 1.
\end{equation}

The top panel of Figure \ref{hr} shows an excess in $H_s(r)$ over
the random sample $H_r(r)$ for small separations. The excess indicates
that the sources in Ophiuchus are not randomly distributed within the
cloud, but clustered on small scales. The middle panel shows that the 
two-point correlation function for the Ophiuchus data 
exceeds zero  by 2.5 $\sigma$ for $r < 4 \times 10^4$ AU, but the random 
distribution shows no correlation ($w(r) = 0$).  The bottom panel of the
figure shows that the correlation function can be fit with a power law, $w
\propto r^{-\gamma}$; the best fit gives $\gamma = 1.5\pm 0.3$ ($\chisq = 1.2$)
for $1 \times 10^4$ AU $< r < 4 \times 10^4$ AU.  
The correlation function for Perseus was characterized by 
$\gamma = 1.25\pm 0.06$ ($\chisq = 0.7$) for 
$2 \times 10^4$~AU~$<r< 2 \times 10^5$~AU.

\citet{stanke05} found $\gamma = 0.63$ out to $r \sim 1\ee5$ AU. \citet{joh00}
also fitted the correlation for the Ophiuchus cluster with a shallower power
law, $\gamma = 0.75$ for $r < 3 \times 10^4$ in the L1688 cluster 
region of Ophiuchus. This power law is
also shown in Figure \ref{hr}, but it clearly does not fit our data.
\citet{joh00} were able to measure the correlation function to
smaller scales, $r = 4.5 \times 10^3$ AU, than this study, which may
result in some discrepancy in the best-fit power law between the two
data sets.  
The correlation function does appear flatter for smaller separations, 
but the slope may be complicated by blending for small separations.
If the correlation function is restricted to sources in the L1688 
cluster, the slope becomes more consistent with those found
by \citet{joh00} and \citet{stanke05}.

We conclude from this analysis that the sources in
Ophiuchus are clearly clustered.
Determining the parameters of the correlation function is complicated
by effects of map size and resolution.

\subsection{Extinction threshold}\label{threshold}

\citet{joh04} suggested that there is a threshold at $A_V = 15$ mag in
Ophiuchus for the formation of cores, with 94\% of the mass in cores
found at or above that extinction level. They did see cores below that
level, but they were faint (low peak flux) and low in mass (low total flux).
They mapped 4 square degrees of
Ophiuchus at 850 \micron\ and compared their data to an extinction map
of Ophiuchus created from 2MASS and R-band data as part of the
COMPLETE project. Comparison of our own extinction map (Fig. \ref{avmap})
with the COMPLETE extinction map shows reasonable agreement, so we 
use our extinction map. 

We have used a simple analysis, comparable to that of \citet{hatchell05}, 
to study the extinction threshold.
Figure~\ref{avprob} plots the probability of finding a 1.1~mm core in 
Ophiuchus as a function of $A_V$.  The probability for a given $A_V$ is 
calculated from the extinction map as the number of 50\arcsec\ 
pixels containing a 1.1~mm core divided by the total number of pixels at 
that $A_V$, and error bars are Poisson statistical errors.

Very few sources are found below $A_V=9$~mag; 93\% of the mass in cores
is found above $\av = 8$ mag (see Table \ref{tab3}).
Thus we suggest that 
$\av = 9$ mag is the extinction limit for finding 1.1~mm cores in Ophiuchus.  
The probability of finding a core increases with $A_V$ beyond this point, 
although the uncertainties are large at high $A_V$ because there are few 
pixels in the extinction map at very high extinctions.  
The probability distribution for Perseus from Paper~I is also shown 
for comparison.  The difference is quite striking; Perseus seems to have
a much lower extinction threshold than does Ophiuchus, even as measured
by us, and still lower than the threshold for Ophiuchus 
found by \citet{joh04}.

To explore the issues further, we plot in Figure \ref{avplots} 
total flux density, peak flux density, 
radius, and mass for $T_D=10$~K  versus $A_V$.  
In contrast to the \citet{joh04} study, we find many (12 out of the total
core sample of 44) bright (total flux density  $>$ 3 Jy) and massive
($M$(10 K) $>$ 2 \msun) sources at $A_V <$ 15 mag.  
Conclusions about thresholds depend on sensitivity to large structures,
slight differences in extinction contours, and differing resolution.
For example, the tentative detections listed in Table \ref{sourcetab} are
in regions with $\av < 9$.

The areas and cloud masses, measured from the extinction, 
above contours of $\av$ are given in Table \ref{tab3}, along with
the masses of cores above the same contours. The percentages of the
total cloud and core masses are also given. Finally, the fraction of the
cloud mass that is found in dense cores, measured for the same
contour level, is given in the last column. This is similar to Table
2 of \citet{joh04}, except that our cloud and core masses are
cumulative and we use bins of $\av = 2$ mag. Even with our lower
threshold, nearly half the total core mass lies above the $\av = 14$
mag contour, which occupies only 2.3\% of the cloud area and 10\% of 
the cloud mass. The dense cores are clearly concentrated in the regions
of high extinction. The ratio of core to cloud mass increases from about
2\% at the lowest contour ($\av = 2$) to an average of 7.4\% for contours
between 8 and 18 mag. (The contour above 20 mag has such little area that
the core mass fraction is not very reliable.)

\section{Summary}

\begin{figure}
\plotone{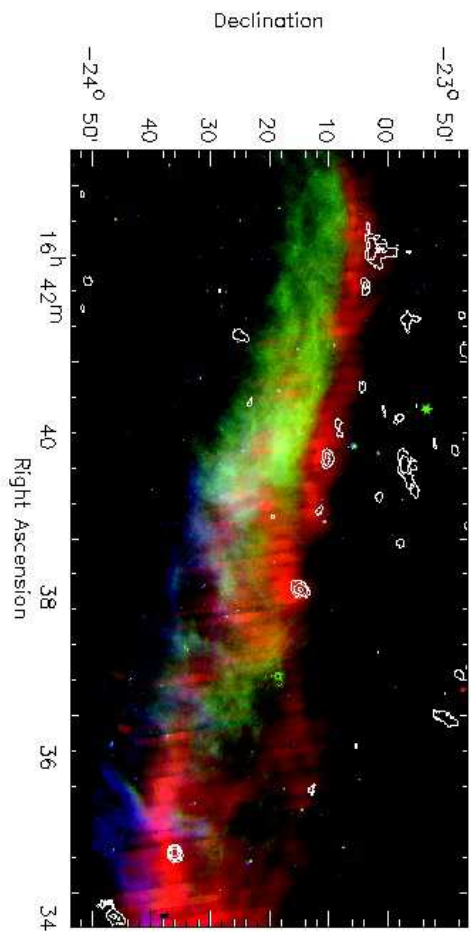} \figcaption{\label{sst3} 
Emission at 1.1 mm seen with Bolocam in the eastern streamer 
(2, 3, 4, 6, 8 $\sigma$ contours) is
overlaid on a three color image from \spitzer, with
IRAC band 4 (8 \micron) in blue, MIPS band 1 (24 \micron) in green,
and MIPS band 3 (160 \micron) in red. The 160 \micron\ map is incompletely
sampled and saturated emission produces stripes. The 160 \micron\ map
has been smoothed, but artifacts remain.  The 1.1 mm emission
does line up with the relatively opaque part of the streamer, as indicated
by weak emission at 8 \micron\ and strong emission at 160 \micron.
}
\end{figure}

\begin{figure}
\plotone{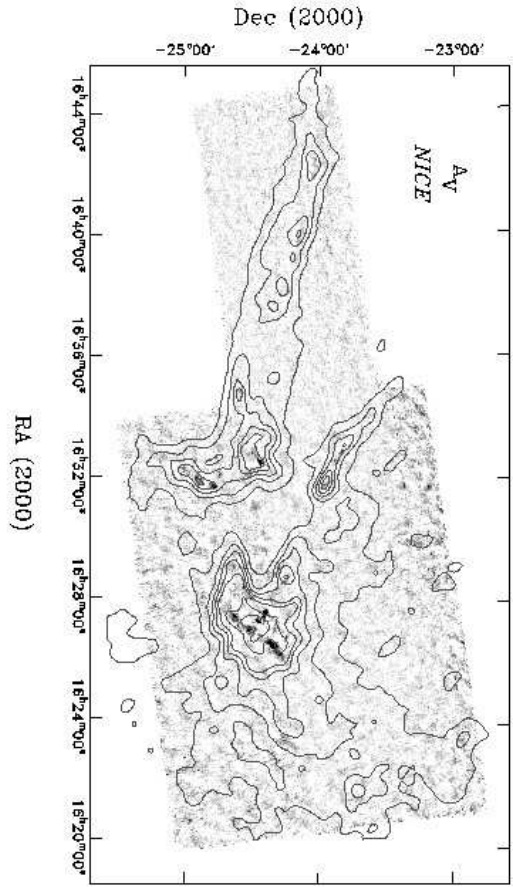} \figcaption{\label{avmap} 
Visual extinction ($A_V$) contours calculated from 2MASS data using 
the ${\it{NICE}}$ method, overlaid on the grayscale 1.1~mm map.  
Contours are $A_V=2$, 4, 6, 8, 10, 15, and 20~mag with an 
effective resolution of $5\arcmin$.
}
\end{figure}

\begin{figure}
\plotone{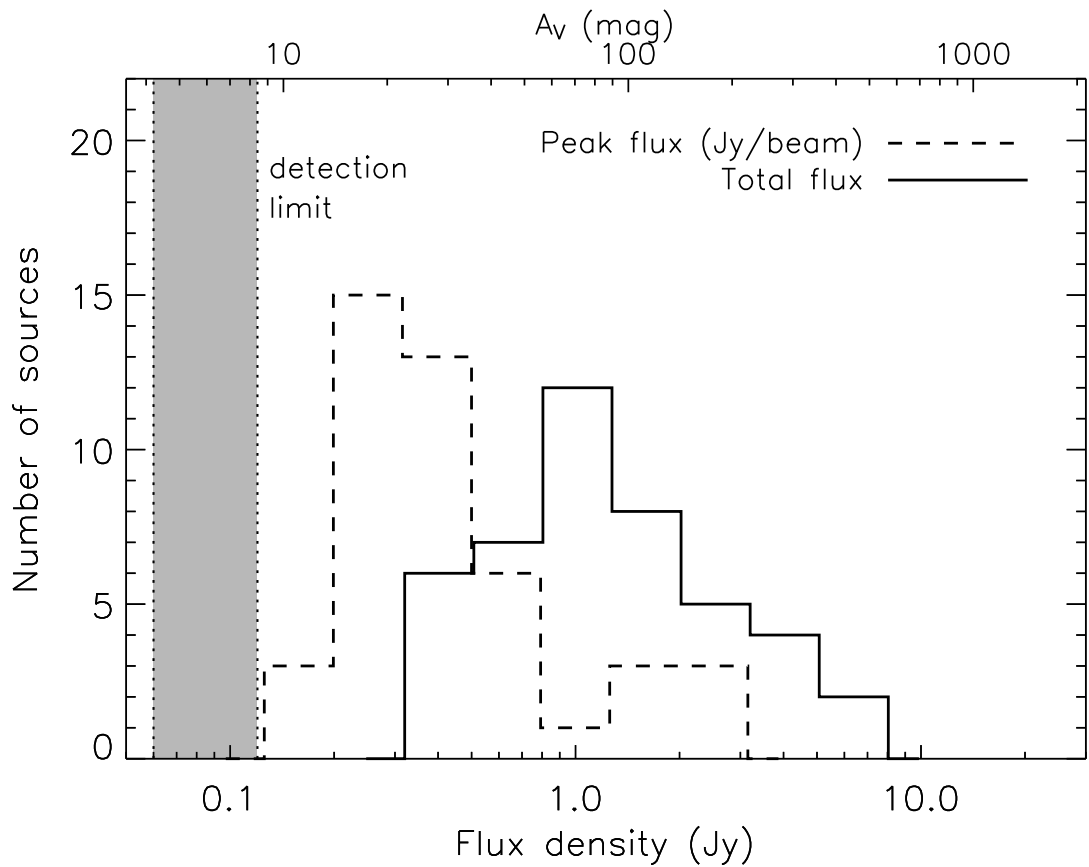} \figcaption{\label{flux} The
distribution of the peak flux densities (dashed line) and total flux
densities (solid line) of the 4-$\sigma$ sources. 
The peak flux density is the peak 
pixel value in the map, in Jy beam$^{-1}$.  
The top axis shows the value of $\av$ inferred from the emission, using
equation 2.
The mean peak flux density of 
the sample is $0.6$~Jy/beam and the mean total flux is $1.6$~Jy.  
The 4-$\sigma$ detection limit varies from $0.06$ to $0.12$~Jy/beam 
across the map 
due to variations in the local noise, although most sources are detected 
in the higher noise regions.  The range in noise in indicated by the 
shaded region. }
\end{figure}

\begin{figure}
\plotone{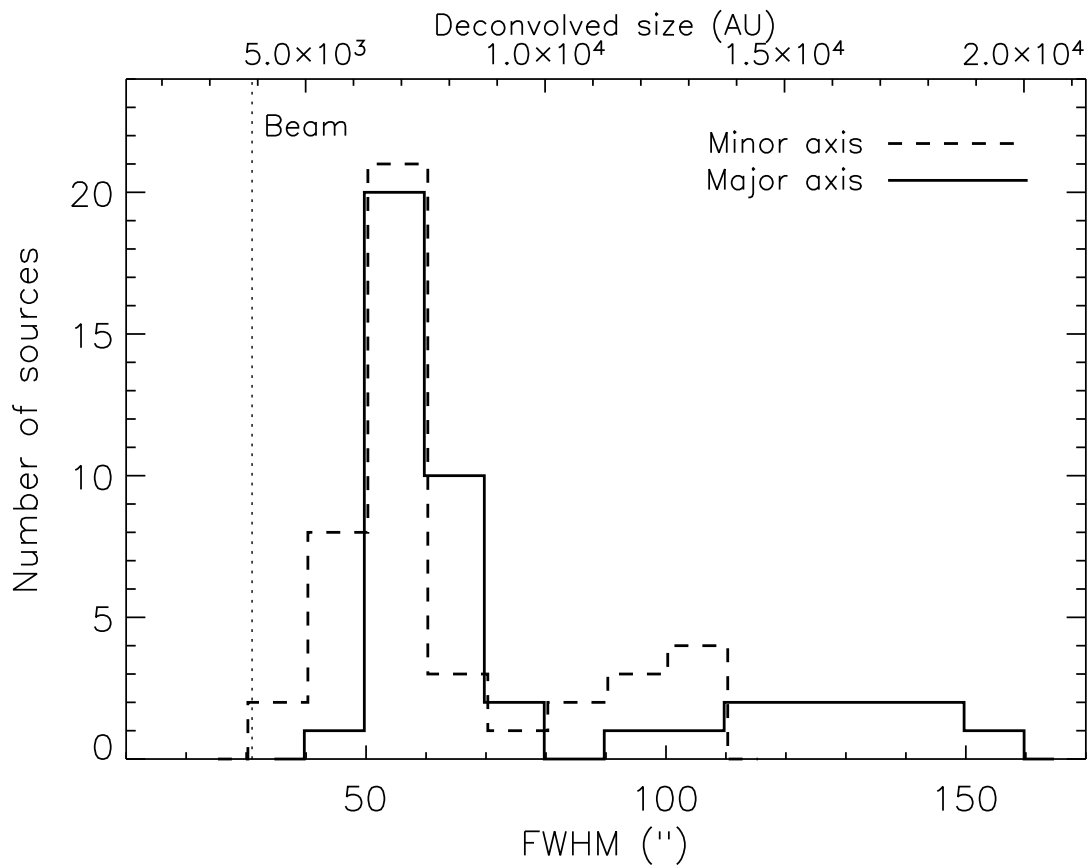} \figcaption{\label{size} The 
distribution of source FWHM  minor axis (dashed line) and 
major axis (solid line), as determined from an elliptical Gaussian
fit. The beam size is indicated by the dotted line. 
The mean FWHM sizes of the sample are $62\arcsec$ (minor) 
and $77\arcsec$ (major).  The top axis gives deconvolved sizes in AU, 
assuming a $31''$ beam and $d=125$~pc.}
\end{figure}

\begin{figure}
\plotone{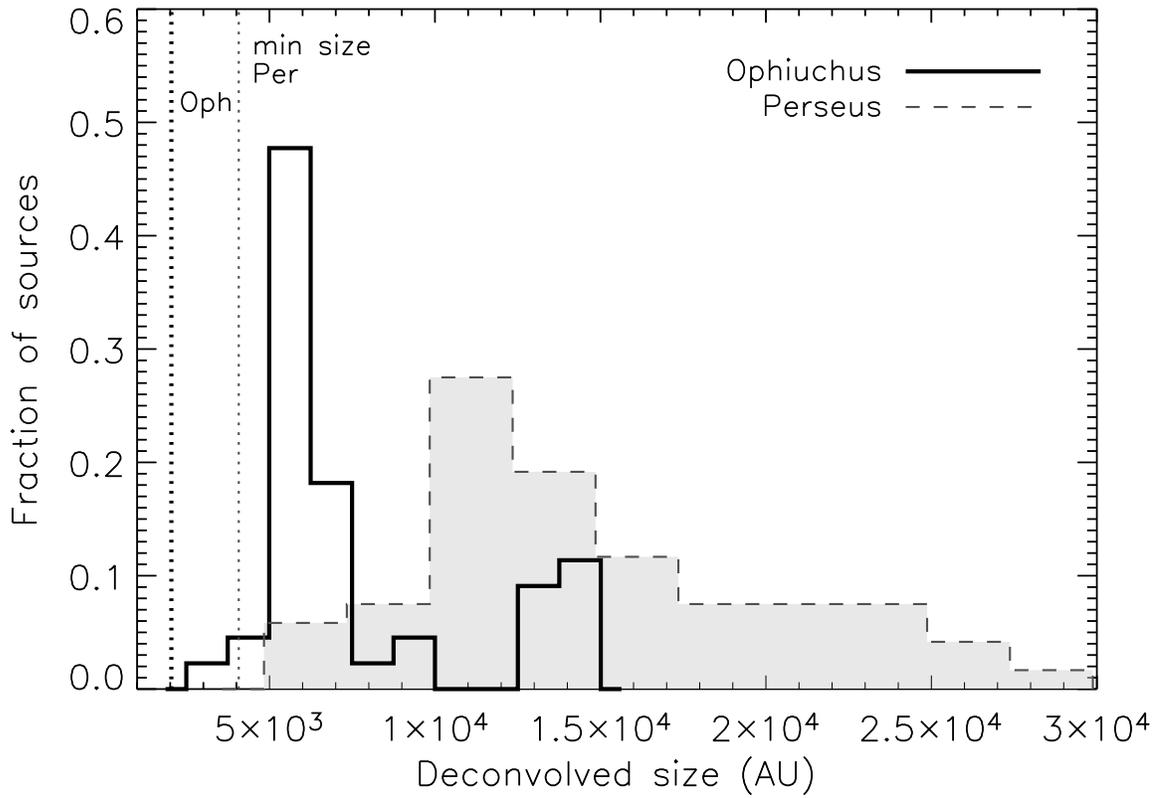} \figcaption{\label{compsize}  
Comparison of the distribution of sizes of sources in Ophiuchus (heavy
line) and Perseus (light line and gray-shaded).  
The fraction of total sources is plotted as a function of deconvolved 
source size in AU.  The vertical dotted lines represent the size of the 
smallest resolvable source (heavy dotted for
Ophiuchus and light dotted for Perseus).  
}
\end{figure}

\begin{figure}
\plotone{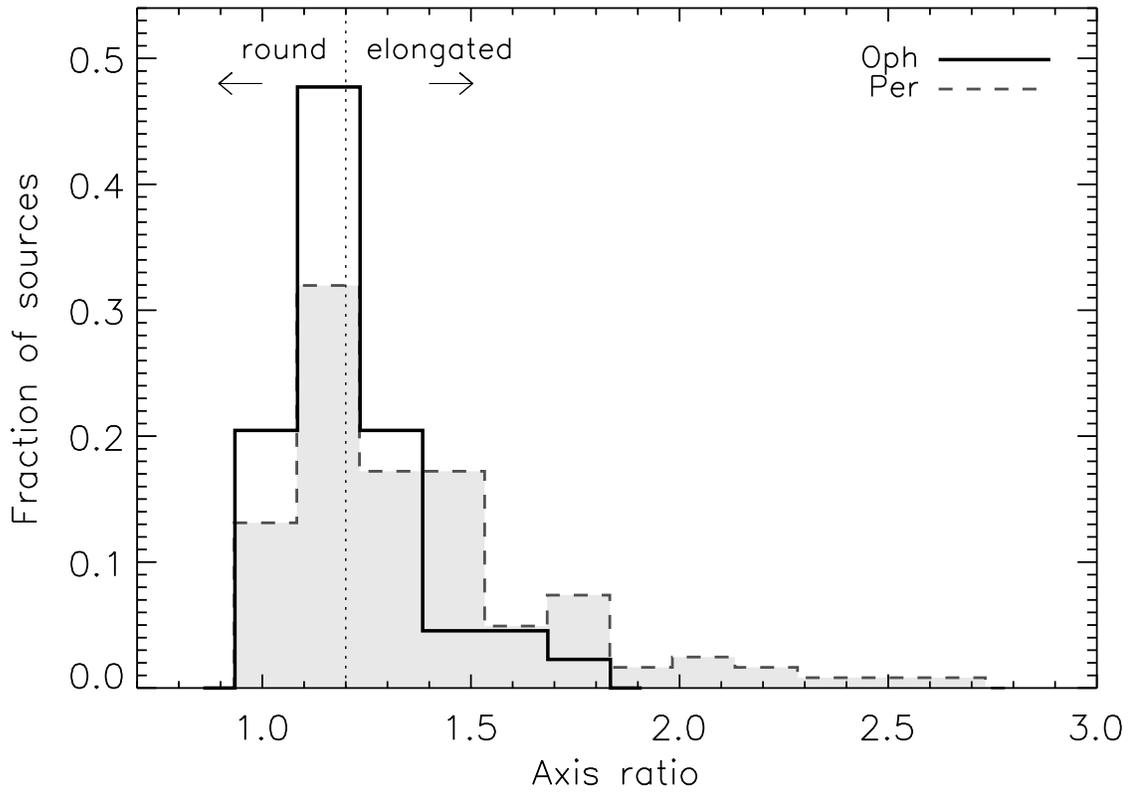} \figcaption{\label{compaxis}  
Comparison of the distribution of axis ratios (measured at the
FWHM) of sources in Ophiuchus and Perseus (light line and gray-shaded).  
Sources with axis ratios $<1.2$ are considered 
round.  There are a larger 
fraction of sources with axis ratios greater than 1.2 in Perseus 
than in Ophiuchus.}
\end{figure}

\begin{figure}
\plotone{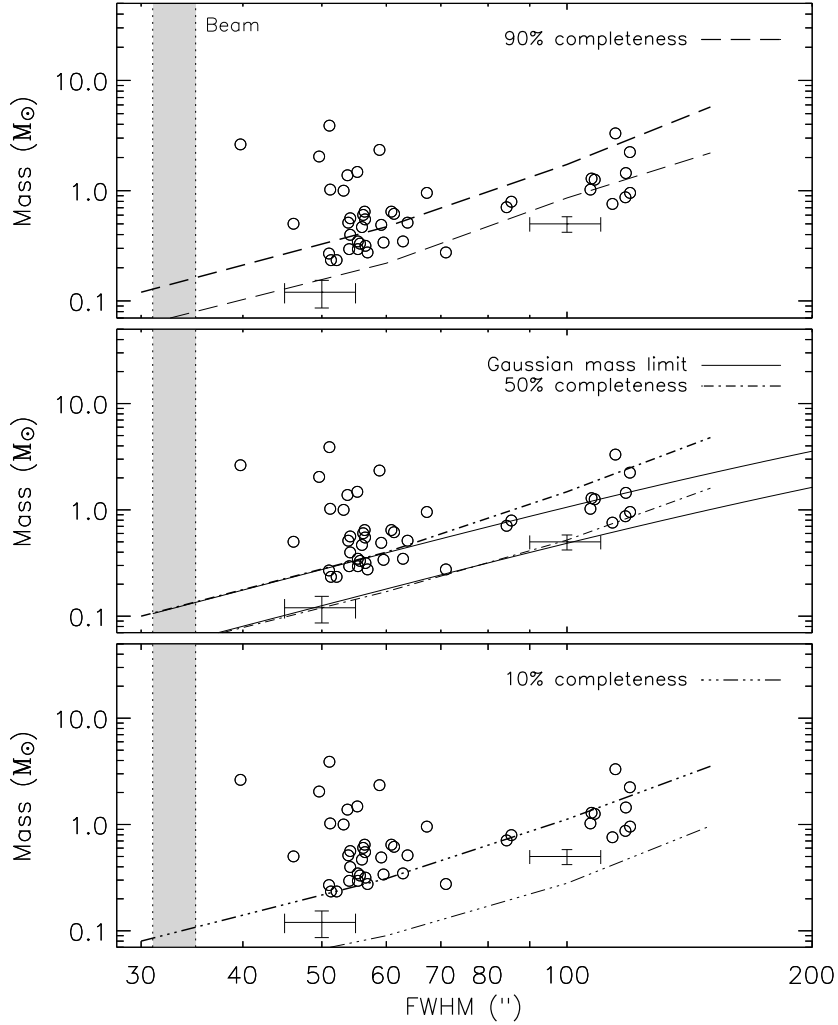} \figcaption{\label{mass_size}
Distribution of source mass ($T_D$ = 10 K) versus FWHM size. The size is
the geometric average of the FWHM of the major and minor axes as given in Table
\ref{sourcetab}. The maximum size of the pointing-smeared 
beam is represented by the shaded regions.  
Solid lines are the 50\% analytic mass detection limit as a 
function of size for Gaussian sources (Eq. 3).  
Empirical 90\%  (top panel), 
50\% (middle panel), and 10\% (bottom panel) completeness limits are 
indicated, derived using Monte Carlo methods with simulated sources and 
taking into account the effects of cleaning, iterative mapping, and 
optimal filtering.  Each completeness limit is calculated both in a 
low rms region (lower line) and a high rms region (upper line).  
Most real sources are found in the higher rms regions.  
Representative error 
bars for $50\arcsec$ and $100\arcsec$ FWHM sources 
near the detection limit are shown, as estimated from the 
results of Monte Carlo simulations.
}
\end{figure}

\begin{figure}
\plotone{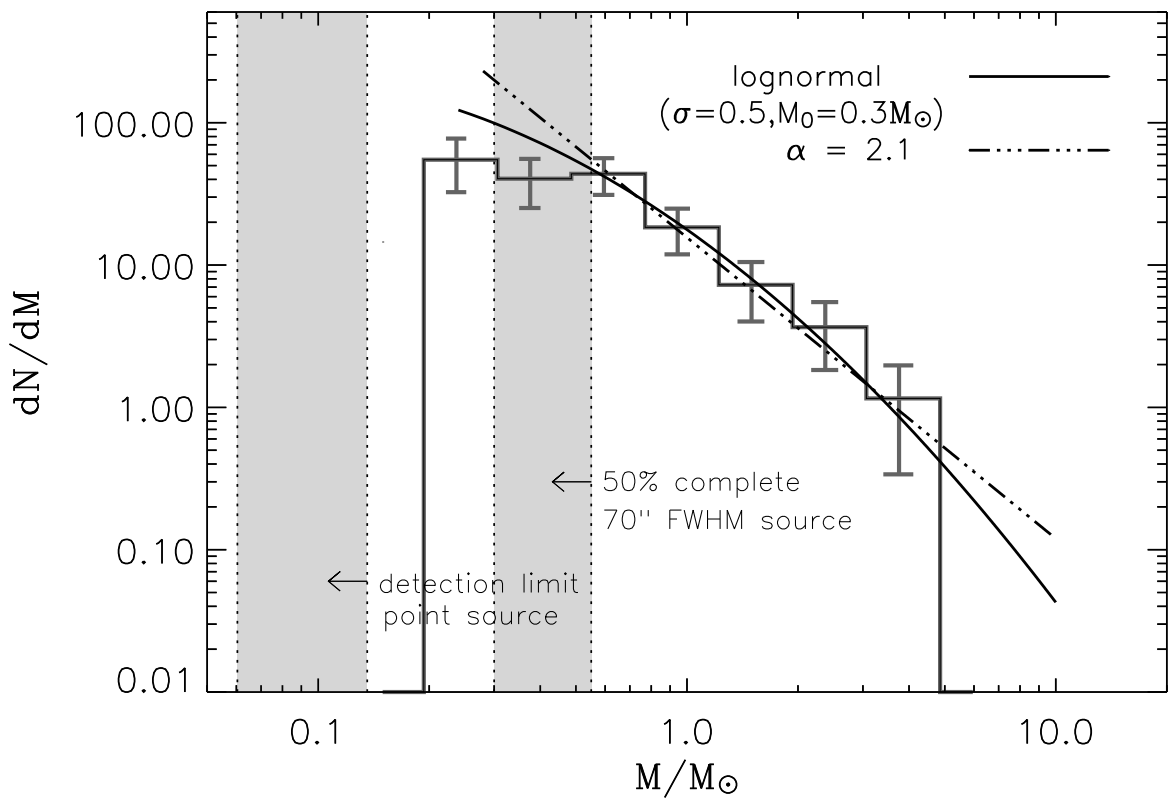} \figcaption{\label{cum_mass} 
Differential mass distribution of all detected sources for 
masses calculated with $T_D = 10$~K. 
The range in completeness, due to the range in 
local rms, is indicated by the shaded regions.  The first is the 
range in detection limit for a point source, and the second is the 
range in 50\% completeness limit for $70\arcsec$ FWHM sources, 
which is similar to the average source size of the sample.
The best fitting power law ($\alpha = 2.1$) is shown,
as well as the best-fitting lognormal function.
}
\end{figure}

\begin{figure}
\plotone{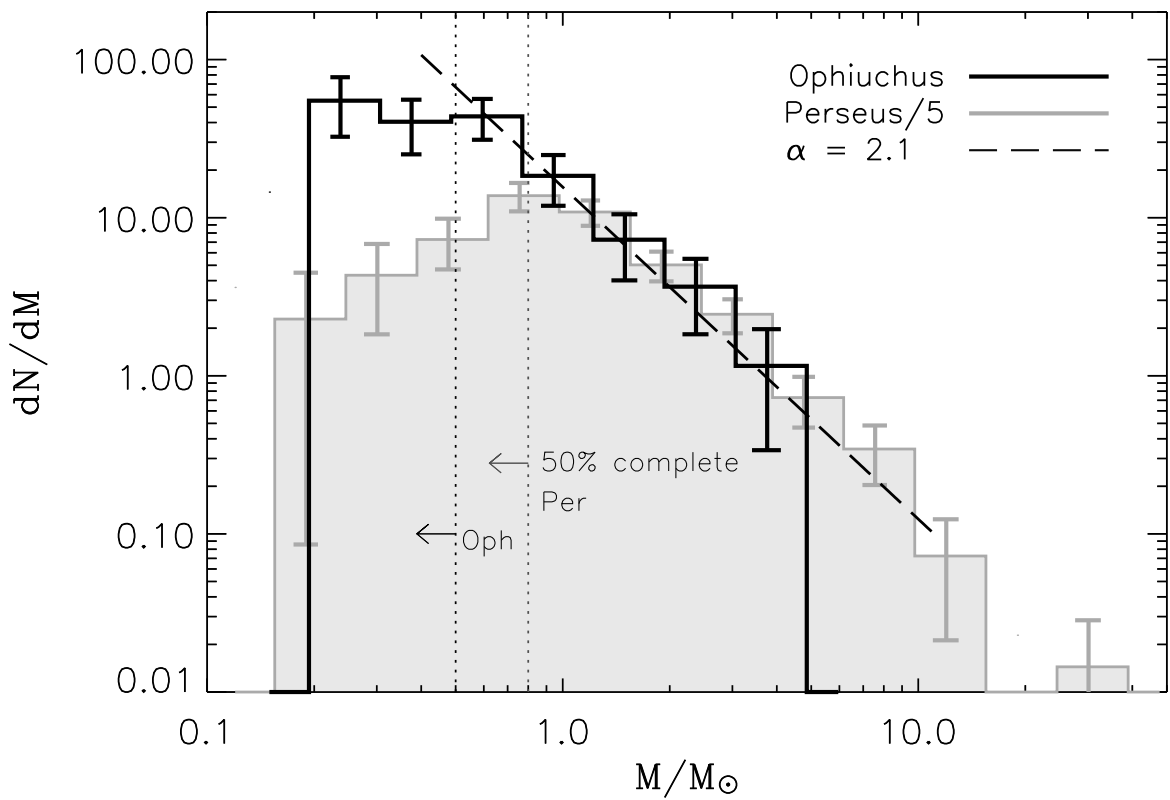} \figcaption{\label{compmass}  
Comparison of the differential mass distributions of sources in Ophiuchus 
and Perseus (light line and gray-shaded).  
The Perseus mass function has been scaled by 1/5 to 
match the amplitude of the Ophiuchus distribution. Uncertainties reflect
only the counting statistics ($\sqrt{N}$).  Vertical dotted
lines show the 50\% completeness thresholds for 70\arcsec\ FWHM sources
for both clouds. The distributions 
appear quite similar in the region where both mass functions are complete. 
}
\end{figure}

\begin{figure}
\plotone{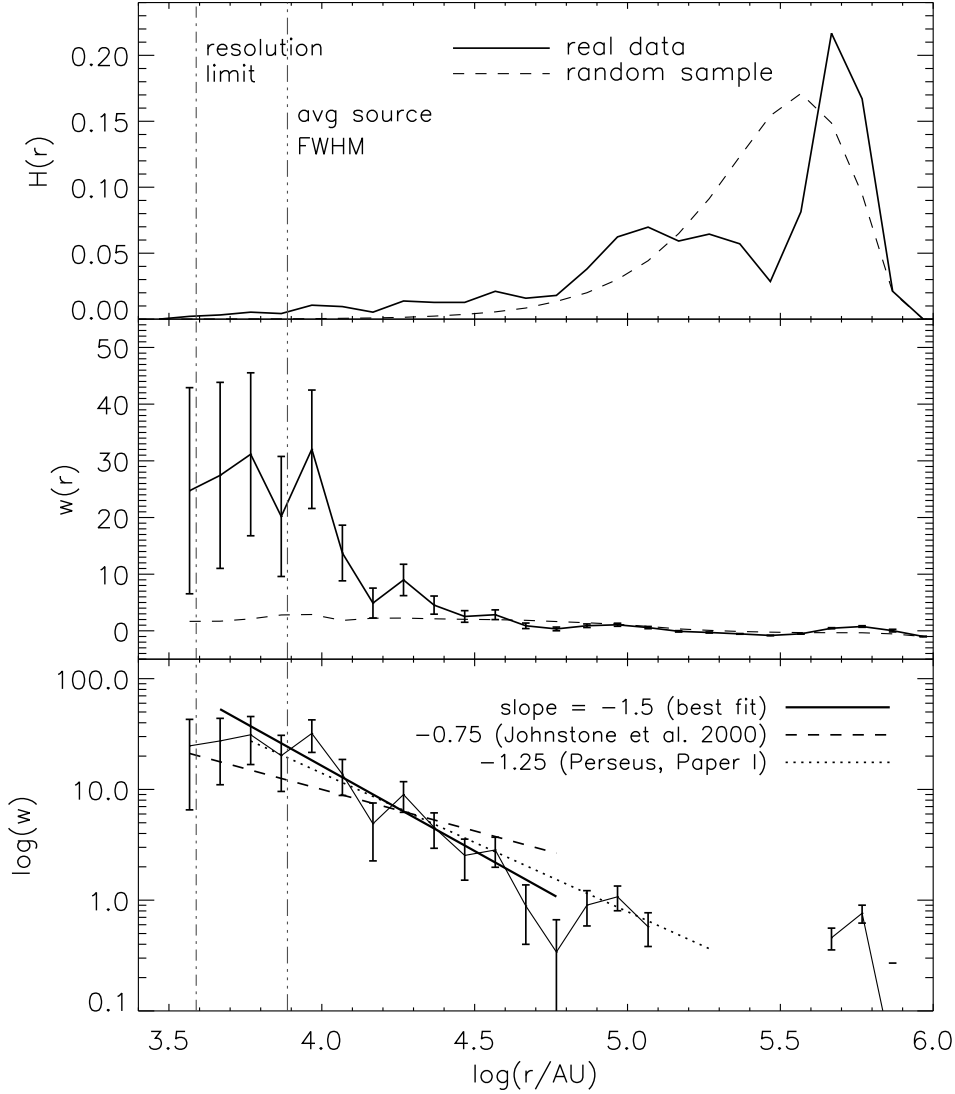} \figcaption{\label{hr} 
Top: $H(r)$, the fractional number of source pairs
between log($r$) and dlog($r$), versus log($r$). 
The solid line indicates the real data, and the dashed line is 
for a uniform random distribution of sources with the same RA/Dec limits 
as the real sample. In all plots, the resolution limit and 
the average source FWHM are shown.
Middle: Two-point correlation function, with $\sqrt{N}$ errors.  
Bottom: Log of the correlation function with power-law fits.  
The best fit slope is $-1.5\pm 0.3$.  The slope found by \citet{joh00} 
in Ophiuchus is shallower ($-0.75$), while the slope found 
in Perseus in Paper~I was similar ($-1.25 \pm 0.06$).  }
\end{figure}

\begin{figure}
\plotone{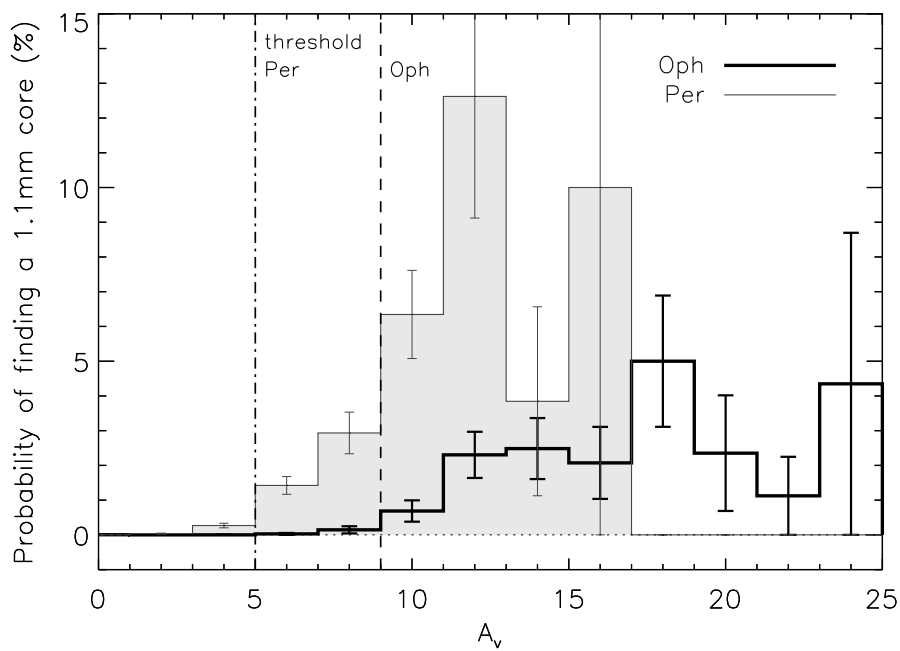} \figcaption{\label{avprob} 
Probability of finding a 1.1~mm core as a function of $A_V$.   
The probability is the number of 50\arcsec\ pixels at a given $A_V$ 
containing one or more 1.1~mm cores, divided by the total number of 
pixels at that $A_V$. Error bars are Poisson statistical errors.  
The dashed vertical line shows our proposed threshold at $A_V = 9$ mag.
The probability distribution for Perseus (light line and gray-shaded) 
from Paper~I is also shown for comparison, with the dash-dotted line showing
the threshold in Perseus.}
\end{figure}

\begin{figure}
\plotone{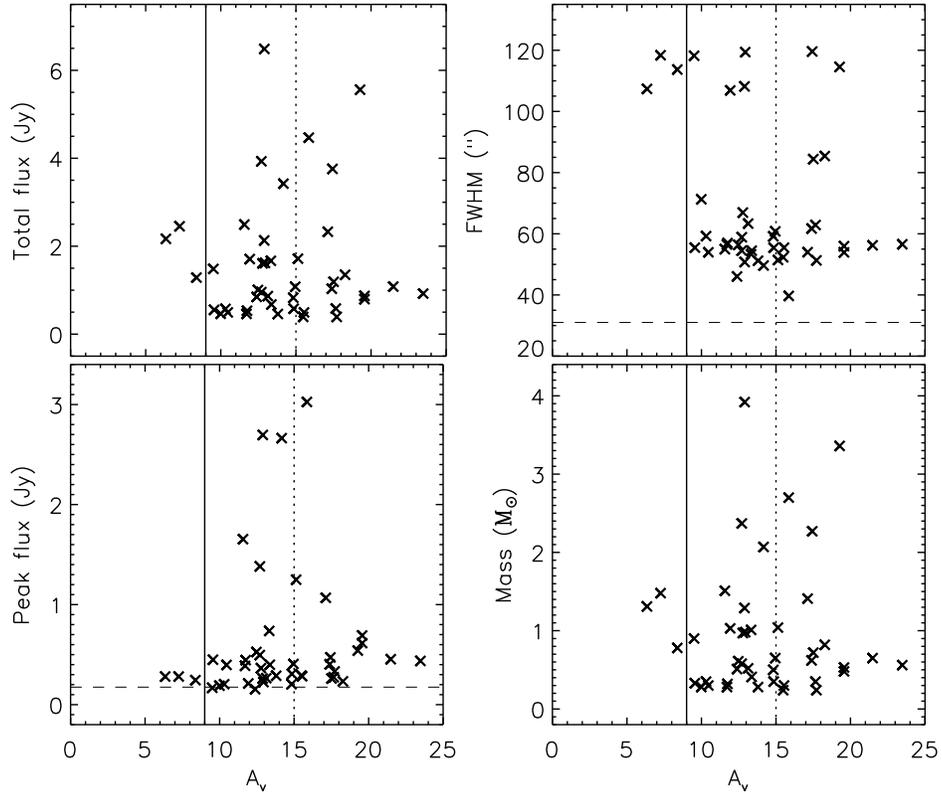} \figcaption{\label{avplots} 1.1~mm source
properties versus $A_V$. 
The dotted vertical lines are the $A_V = 15$ mag threshold
proposed by \citet{joh04}. The solid vertical line is the
$A_V$ = 9 mag extinction threshold from Figure~\ref{avprob}. 
The dashed horizontal lines are the
beam size in the upper right panel and the average 4-$\sigma$ detection
limit in the lower left panel.  
}
\end{figure}

We presented a 1.1~mm dust continuum emission map of 10.8 
deg$^2$ of the Ophiuchus molecular cloud. We detected 44 sources at 4 $\sigma$
or greater, almost all concentrated around well known clusters (near
the dark clouds L1688, L1689, and L1709). 
Some weaker emission (3 $\sigma$) was seen along the eastern streamer 
of the cloud, coincident with a filament seen in both extinction (Fig.
\ref{avmap}) and emission at 160 \micron\ (Fig. \ref{sst3}). 
These cores have been previously seen in maps of CO,
but these are the first millimeter dust continuum
observations of the streamer. We did not detect any emission in the
northeastern streamer.

Visually, the 4-$\sigma$ sources appear highly clustered, and this impression
is confirmed by the two-point correlation function, the fraction of 
multiple sources, and the median separation. Fully  82\% of the sources
are classified as multiple (i.e., another source lies within 3\arcmin).
Most of the cloud area has no detectable sources.

Most sources are round as measured at the FWHM, but many are elongated
when measured at lower contour levels. This difference probably reflects
the fact that many are relatively spherical condensations within filaments.
Filamentary structure with condensations along the filaments is the
dominant morphological theme.

The total mass of the sources is only 42 \msun, about 0.4 to 1.8\% of the total
cloud mass, lower than in Perseus (Paper I), while the total mass corresponding
to emission above 4 $\sigma$ is 79 \msun. The differential mass distribution
can be fitted with a power law with slope $-2.1\pm 0.3$ or a lognormal function.
It is similar to that in Perseus, but does not extend to as high a mass, 
with the most massive core containing only 3.9 \msun. The mean densities
are quite high, averaging $1.6\ee6$ \cmv, implying a short free-fall time.

Millimeter continuum sources are seen for $A_V$ above a threshold value
of 9 mag, higher than in Perseus, but lower than found in previous studies
of Ophiuchus by \citet{joh04}. About half the total mass of dense cores
are in contours of extinction below $A_V = 14$ mag, near the threshold
seen by \citet{joh04}. Still, the cores are clearly concentrated in
a small fraction of the cloud area and mass, and in regions of relatively
high extinction.

Future analysis of these data in combination with the c2d \spitzer\ 
maps of Ophiuchus
will give a more complete picture of star formation in the cloud. 
Additionally, the environments of other clouds in the c2d survey will 
be compared using the combined infrared and millimeter data sets in 
future work.

\acknowledgments

We would like to thank A. Urban for assistance observing at the CSO, and
other members of the larger Bolocam team for instrumental support, including
A. Goldin, A. Lange, P. Maloney, and P. Rossinot.  
We thank the Lorentz Center in Leiden for hosting several meetings
that contributed to this paper.
Support for this work, part of the
\spitzer\ Legacy Science Program, was provided by NASA through contracts
1224608 and 1230782 issued by the Jet Propulsion Laboratory,
California Institute of Technology, under NASA contract 1407. 
Bolocam was built and commissioned under grants NSF/AST-9618798 and
NSF/AST-0098737.
KEY and GL were supported by NASA under Grants NGT5-50401 and
NGT5-50384, respectively, issued through the Office of Space Science. 
Additional support came from NASA Origins grant NNG04GG24G to NJE
and NSF grant AST 02-06158 to JG.
MLE acknowledges support of an NSF Graduate Research Fellowship.
SG was supported in part by a Millikan fellowship at Caltech 
and CSO grant NSF/AST-9980846.


\clearpage
\ProvidesFile{table1_new.tex}
\begin{deluxetable}{lccccc}
\tabletypesize{\scriptsize}
\tablecolumns{6}
\tablewidth{0pc}
\tablecaption{Sources Found in Ophiuchus \label{sourcetab}}
\tablehead{
\colhead{ID} & \colhead{RA (2000)} & \colhead{Dec (2000)} & \colhead{Peak} &  \colhead{S/N}  & \colhead{other names}  \\ 
 \colhead{} & \colhead{($h ~~m ~~s$)} & \colhead{($\degr~~ \arcmin~~ \arcsec$)} & \colhead{(Jy/beam)} &  \colhead{}  & \colhead{}   
}
\startdata
Bolo 1  &  16 25 59.1 &  -24 18 16.2  &  0.26 (0.03)  &  4.5  &       \\
Bolo 2  &  16 26 08.1 &  -24 20 00.6  &  0.39 (0.03)  &  4.7  & CRBR 2305.4-1241  ?      \\
Bolo 3  &  16 26 09.6 &  -24 19 15.6  &  0.31 (0.03)  &  4.1  &  SMM J16261-2419 (1)     \\
Bolo 4  &  16 26 09.9 &  -24 20 28.6  &  0.40 (0.03)  &  6.2  & GSS26?     \\
Bolo 5  &  16 26 20.7 &  -24 22 17.0  &  0.37 (0.04)  &  4.3  & GSS30-IRS3 (2); SMM J16263-2422 (1); LFAM1 (3)       \\
Bolo 6  &  16 26 22.9 &  -24 20 00.9  &  0.27 (0.03)  &  4.5  & SMM J16263-2419 (1)      \\
Bolo 7  &  16 26 24.7 &  -24 21 07.5  &  0.41 (0.04)  &  4.9  &  A-MM4? (5)    \\
Bolo 8  &  16 26 27.2 &  -24 22 26.7  &  1.38 (0.04)   &  16.0  & SM1 FIR1 (4); A-MM5/6?? (5); SMM J16264-2422 (1)     \\
Bolo 9  &  16 26 27.6 &  -24 23 36.6  &  2.70 (0.04)   &  45.2  & SM1 FIR2 (4); SM1N      \\
Bolo 10  &  16 26 29.7 &  -24 24 28.8  & 2.66 (0.04)    &  47.5  & SM2      \\
Bolo 11  &  16 26 32.6 &  -24 24 45.3  & 1.25 (0.04)    &  14.8  &  A-MM8 (5)     \\
Bolo 12  &  16 27 00.7 &  -24 34 17.0  & 0.54 (0.03)   &  8.4  & SMM J16269-2434 (1); C-MM3 (5)     \\
Bolo 13  &  16 27 04.3 &  -24 38 47.4  & 0.23 (0.03)   &  4.8  & E-MM2d (5); SMM J16270-2439 (1)      \\
Bolo 14  &  16 27 07.9 &  -24 36 54.3  & 0.26 (0.03)   &  4.4  & SMM J16271-2437a/b (1); Elias29?      \\
Bolo 15  &  16 27 12.2 &  -24 29 18.9  & 0.44 (0.04)   &  6.1  & B1-MM2/3 (5); IRAS 16242-2422; SMM J16272-2429 (1)      \\
Bolo 16  &  16 27 15.1 &  -24 30 12.6  & 0.45 (0.04)   &  7.1  &  SMM J16272-2430 (1); B1-MM4 (5)    \\
Bolo 17  &  16 27 22.3 &  -24 27 36.3  & 0.40 (0.04)   &  4.4  &  B2-MM4 (5)     \\
Bolo 18  &  16 27 25.2 &  -24 40 28.9  & 0.47 (0.03)   &  9.7  &  F-MM2 (5); IRS43? (5)     \\
Bolo 19  &  16 27 27.0 &  -24 26 57.1  & 0.62 (0.04)   &  7.4  &  SMM J16274-2427s (1)     \\
Bolo 20  &  16 27 29.1 &  -24 27 11.1  & 0.69 (0.04)   &  9.4  & B2-MM8 (5); SMM J16274-2427b (1)      \\
Bolo 21  &  16 27 33.1 &  -24 26 48.8  & 0.53 (0.04)   &  6.8  & B2-MM15 (5); SMM J16275-2426 (1)       \\
Bolo 22  &  16 27 33.4 &  -24 25 57.3  & 0.50 (0.03)   &  9.4  & B2-MM13 (5); SMM J16275-2426 (1)      \\
Bolo 23  &  16 27 36.7 &  -24 26 36.2  & 0.40 (0.03)   &  5.1  & B2-MM17 (5)      \\
Bolo 24  &  16 27 58.3 &  -24 33 09.7  & 0.29 (0.03)   &  6.0  &       \\
Bolo 25  &  16 28 00.1 &  -24 33 42.8  & 0.33 (0.03)   &  6.0  & H-MM1 (8)      \\
Bolo 26  &  16 28 21.0 &  -24 36 00.0  & 0.23 (0.03)   &  5.5  &      \\
Bolo 27  &  16 28 32.1 &  -24 17 43.4  & 0.17 (0.03)   &  4.1  &  D-MM3/4 (5)     \\
Bolo 28  &  16 28 57.7 &  -24 20 33.7  & 0.28 (0.03)   &  5.8  &  I-MM1 (8)    \\
Bolo 29  &  16 31 36.4 &  -24 00 41.7  & 0.28 (0.03)   &  7.0  &  L1709-SMM1 (6); IRS63     \\
Bolo 30  &  16 31 37.2 &  -24 01 51.9  & 0.29 (0.03)   &  6.3  &  L1709-SMM3,5 (6)     \\
Bolo 31  &  16 31 40.0 &  -24 49 58.0  & 0.45 (0.03)   &  6.8  &       \\
Bolo 32  &  16 31 40.4 &  -24 49 26.4  & 0.45 (0.03)   &  7.4  &       \\
Bolo 33  &  16 31 52.6 &  -24 58 00.8  & 0.20 (0.03)   &  4.5  &       \\
Bolo 34  &  16 31 58.0 &  -24 57 38.8  & 0.26 (0.03)   &  5.2  &       \\
Bolo 35  &  16 32 00.6 &  -24 56 13.9  & 0.19 (0.03)   &  4.5  & IRAS 16289-2450; L1689S; IRS67       \\
Bolo 36  &  16 32 22.5 &  -24 27 47.5  & 1.65 (0.03)    &  17.9  &       \\
Bolo 37  &  16 32 24.7 &  -24 28 51.2  & 3.03 (0.03)    &  80.7  & IRAS 16293-2422      \\
Bolo 38  &  16 32 28.6 &  -24 28 37.5  & 1.07 (0.03)    &  17.0  &       \\
Bolo 39  &  16 32 30.1 &  -23 55 18.4  & 0.25 (0.03)   &  4.8  & L1709-SMM2 (6)      \\
Bolo 40  &  16 32 30.8 &  -24 29 28.3  & 0.74 (0.03)   &  14.7  &       \\
Bolo 41  &  16 32 42.3 &  -24 31 13.0  & 0.15 (0.02)   &  4.5  &       \\
Bolo 42  &  16 32 44.1 &  -24 33 21.6  & 0.20 (0.03)   &  4.3  &       \\
Bolo 43  &  16 32 49.2 &  -23 52 33.9  & 0.28 (0.03)   &  4.1  & L1709-SMM4? (6); LM182     \\
Bolo 44  &  16 34 48.3 &  -24 37 24.6  & 0.21 (0.02)   &  5.6  & L1689B-3      \\
\cutinhead{Tentative detections in the eastern streamer}
Bolo 45  &  16 38 07.8 &  -24 16 36.4  & 0.15 (0.02)   &  3.6  & $\rho$-Oph~10 (7)      \\
Bolo 46  &  16 39 15.8 &  -24 12 20.8  & 0.10 (0.02)   &  3.2  &       \\
Bolo 47  &  16 41 44.5 &  -24 05 20.4  & 0.14 (0.02)   &  3.0  &       \\
Bolo 48  &  16 41 55.6 &  -24 05 41.6  & 0.12 (0.02)   &  3.0  &       \\

\enddata
\tablecomments{Numbers in parentheses are $1\sigma$ errors. 
The peak flux density is the peak pixel value in the $10\arcsec$ pixel$^{-1}$
unfiltered map (without the Wiener filter applied).
The uncertainty in the peak flux density is the local (calculated
within a 400\arcsec\ box) rms beam$^{-1}$, calculated
from the noise map, and does not include an additional 15\% systematic
uncertainty from calibration uncertainties and residual errors after
iterative mapping.
 Other names listed are the most common identifications from the literature, 
and are not meant to be a complete list.  References -- (1) Johnstone 
et al. 2000; (2) Castelaz et al. 1985, Weintraub et al. 1993; (3) Leous 
et al. 1991; (4) Mezger et al. 1992; (5) Motte, Andre \& Neri 1998; 
(6) \citet{visser02}; (7) \citet{tac00}; (8) \citet{joh04}
}
\end{deluxetable}

\clearpage
\pagestyle{empty}
\input{table2}
\clearpage
\pagestyle{plaintop}
\input{table3}			
\clearpage

\epsscale{0.8}

\epsscale{0.8}

\epsscale{0.9}

\epsscale{0.7}
\epsscale{1}

\epsscale{0.7}
\epsscale{1}

\epsscale{0.7}
\epsscale{1}

\epsscale{0.8}

\end{document}